\documentclass[journal,twoside]{asmb}

\begin{document}
\title{Design and Performance Analysis of Hybrid FSO/THz Relay with Aerial RIS for Future NTN-Integrated $6$G Wireless Communications}

\author{
\IEEEauthorblockN{Al Nahian Mugdho, Md. Ibrahim,
\textit{Graduate Student Member, IEEE}, A. S. M. Badrudduza, \textit{ Member, IEEE}, Md. Abdur Rakib, and Imran Shafique Ansari, \textit{Senior Member, IEEE}}}

\twocolumn[
\begin{@twocolumnfalse}
\maketitle
\begin{abstract}
\section*{Abstract}

In the context of emerging sixth-generation ($6$G) wireless networks, reconfigurable intelligent surfaces (RISs) are gaining prominence for their ability to intelligently control electromagnetic wave propagation and enhance backhaul communication performance. In this paper, we propose a novel dual-hop wireless network, where the first hop consists of hybrid free-space optics (FSO) / terahertz (THz) link, and the second hop incorporates an aerial RIS based radio frequency (RF) link. To provide a comprehensive performance evaluation, a comparative analysis of two switching strategies is conducted: (1) hard switching and (2) soft switching. Novel closed-form expressions are derived for key performance metrics, including outage probability and bit error rate. These expressions are then utilized to investigate the impact of various system parameters. Our proposed hybrid model demonstrates a $52.54\%$ performance improvement over traditional RF-FSO framework. Moreover, the integration of an aerial RIS in the second hop enhances system performance by $41.39\%$. Numerical findings suggest that strategically placing the aerial RIS at a lower altitude and maintaining an equal, shorter distance from both communication endpoints significantly improves overall system performance. {\color{black} To analyze the response under high signal-to-noise ratio (SNR) conditions, asymptotic analysis is performed and the diversity order of the system is determined. }Finally, the analytical results are validated through Monte Carlo simulation.
\end{abstract}

\begin{IEEEkeywords}
\section*{Keywords} 

RIS, THz link, FSO, UAV, dual-hop, Rician, Málaga distribution.
\end{IEEEkeywords}
\end{@twocolumnfalse}
]
\section{Introduction}
\subsection{Background}
Reconfigurable Intelligent Surfaces (RISs) have emerged as a promising technology in wireless communication, offering significant improvements spectrum efficiency and energy efficiency \cite{r1}. The incorporation of unmanned aerial vehicle (UAV) mounted RIS units adds another layer of adaptability by allowing dynamic repositioning and optimized signal paths for improved coverage \cite{li2023ris}. At the same time, free-space optics (FSO) and terahertz (THz) communication have emerged as key technologies for next-generation systems. These technologies offer ultra-high bandwidth, low latency, and strong resistance to electromagnetic interference. Utilizing both FSO and THz links concurrently presents a practical solution for achieving reliable communication. To further enhance this model, a dual-hop network with an intermediate relay strengthens the system by extending the communication range and mitigating channel impairments. With proper coordination, a system employing parallel FSO and THz links in the first hop and a RIS-mounted UAV for the RF link in the second hop delivers a highly efficient communication framework capable of addressing environmental obstacles.

\subsection{Literature Study}
With the increasing demand for high-speed data transmission and the growing need for reliable communication, hybrid RF/FSO systems have gained significant research attention \cite{n1}. Recent works have explored multiple aspects of these networks \cite{n2,n3,n22,n23,n4,n5,r11,n6,n7,n8,thz}. In \cite{n2}, the authors proposed a parallel RF/FSO system that incorporated a Gamma-Gamma (GG) turbulence model for the FSO link and Rician fading for the RF link. Another study in \cite{n3} evaluated the performance of a hybrid RF/FSO system where the findings revealed that the hybrid model successfully overcomes the inherent limitations of FSO links, which are particularly vulnerable to misalignment fading and atmospheric variations. An adaptive switching scheme for the hybrid model was introduced in \cite{n4}. The study in \cite{n5} demonstrated a hard-switching mechanism for a parallel FSO/RF system, analyzing the impacts of fading, beam wander, pointing errors, and atmospheric turbulence. A comparative study in \cite{r11} demonstrated that soft switching outperforms hard switching, particularly under low SNR scenario. The research in \cite{n6} examined the performance of a hybrid RF/FSO network. In \cite{n7}, a hybrid model combining optical and RF channels was examined, in which the authors used the Málaga distribution to characterize atmospheric turbulence in the FSO links and used the Nakagami-$m$ distribution to model the fading of the mmWave link. Another work in \cite{thz} proposed a dual-hop hybrid model applicable to non-terrestrial networks (NTN)-integrated scenarios.  

UAV-assisted RF-FSO systems have garnered considerable research interest due to their ability to enhance connectivity and extend coverage, especially for future NTN architectures. Several recent studies have investigated UAV-based communication models \cite{r6, r13, r14, r2, n26, n29, n30, n25, r8, r9}. For instance, the work in \cite{r6} examined the effects of atmospheric turbulence and UAV positional variations during hovering on the performance of FSO systems. A non-orthogonal multiple access (NOMA) based UAV framework was analyzed in \cite{r13}, while \cite{r14} demonstrated that optimizing beam width and positioning of UAVs can significantly improve the system performance. Furthermore, dual-hop relaying has emerged as a promising approach to extend wireless coverage by enabling information to be forwarded via an intermediate node between the source and destination. A substantial body of research has focused on evaluating the performance of dual-hop systems \cite{n17, n9, n10, n16, n11, n18, n19, n12, n13, n20, n14, n21, n15}. In \cite{n9}, a dual-hop FSO-RF system was studied, where the RF link was modeled using the Nakagami-$m$ distribution and the FSO link using Gamma-Gamma (GG) atmospheric distributions. The authors derived closed-form expressions (CFEs) for key performance metrics such as outage probability (OP), ergodic capacity, and average bit error rate (ABER). The study in \cite{n10} analyzed a dual-hop system utilizing both intensity modulation with direct detection (IM/DD) and heterodyne detection (HD). A comparative analysis of HD and IM/DD techniques in terms of OP was presented in \cite{n16}. Atmospheric turbulence and fading effects were addressed in \cite{n11}, while an amplify-and-forward (AF)-based mixed FSO-RF system was introduced in \cite{n12}.

Recently, several studies in \cite{r15, r17, r18,odeyemi2022performance,vishwakarma2024ris,sharma2022performance,r9,ris,mahmoud2021intelligent} have investigated the performance of RIS-assisted wireless models. The authors in \cite{r17} presented a comparative analysis between RIS-equipped and RIS-aided source scenarios for a dual-hop network. Another work in \cite{r15} demonstrated that a higher number of reflecting elements notably enhances overall system performance. The works in \cite{odeyemi2022performance,vishwakarma2024ris,sharma2022performance} analyzed the performance of RIS-based hybrid networks. Despite their potential, UAVs struggle to meet the increasing demand for high data rates due to limitations in size, power, fuel efficiency, and network disruptions. To overcome these issues, the incorporation of RIS into UAV systems has transformed wireless networking by offering energy-efficient solutions \cite{gong2020toward}. 
Only a limited number of studies \cite{ris,mahmoud2021intelligent} have performed performance analysis of RIS-assisted UAV communication systems. The authors in \cite{ris} revealed that deploying RIS-assisted UAVs can significantly improve network coverage and increase channel capacity. The study in \cite{mahmoud2021intelligent} demonstrated that the incorporation of an RIS-assisted UAV introduces improved spectral efficiency of IoT (Internet of Things) networks.

\subsection{Motivations and Contributions}
The existing literature shows that current research efforts
are primarily focused on RF and FSO networks due to the
increasing demands of next-generation $6$G wireless communication. FSO links are well regarded for enabling high-capacity transmission in ultra-dense network environments. However, their performance is significantly hindered by factors such as pointing errors and the requirement for a clear line-of-sight. In contrast, RF links provide more consistent performance under adverse weather but still suffer from fading [28]. To address these limitations, most of the works have introduced parallel FSO/RF systems in recent times \cite{n2,n3,n23,n4,n5}. However, due to the superior bandwidth and data rate potential of THz links over traditional RF links, the proposed model incorporates a hybrid FSO/THz backhaul network to improve overall system performance. Although hybrid networks have been extensively explored, there is a notable gap in the analysis of dual-hop hybrid models. While most existing studies on hybrid systems emphasize hard switching techniques \cite{n4,n5,r11}, the performance of soft switching mechanisms in dual-hop hybrid networks remains unexplored. This work introduces a novel dual-hop hybrid relaying framework that incorporates both hard and soft switching approaches and provides a comparative
analysis. Although UAV-based communication systems have
been extensively investigated \cite{r2, n26, n29, n30, n25}, the integration of aerial RIS into dual-hop models remains a relatively novel concept in the existing literature. Therefore, the RIS-enabled UAV communication on dual-hop network presents a promising avenue for further investigation. To the best of authors knowledge, the performance assessment of a hybrid FSO/THz dual-hop realying network employing a RIS-mounted aerial platform remains an open research challenge.

\begin{table*}[!ht]
\centering
\caption{Overview of Existing Works}
\scriptsize 
\renewcommand{\arraystretch}{1.5} 
\setlength{\tabcolsep}{2pt} 
\begin{tabular}{|c|c|c|c|c|c|}
\hline
\textbf{Ref.} & \textbf{Channel Model} & \textbf{Natwork Configuration} & \textbf{Switching Technique} & \textbf{Aerial RIS} & \textbf{Performance Metric} \\ \hline
\cite{ris} & RF (RIS)  & Single hop & -- &\checkmark & OP, BER \\ 
\cite{mahmoud2021intelligent} & RF (RIS)  & Single hop & -- &\checkmark & OP, SER, EC \\ 
\cite{vishwakarma2024ris} & FSO/THz (RIS)   & Hybrid & Hard switching & -- & OP, SER \\ 
\cite{sharma2022performance} & FSO/RF (RIS)   & Hybrid & Hard switching & -- & OP, BER, EC \\
\cite{r9}  & FSO--RF (RIS)  & Dual-hop & -- & -- & OP, BER, ACC \\
\cite{r17}  & RF (RIS)--FSO  & Dual-hop & -- & -- & OP, ASEP \\ 
\cite{r18} & RF (RIS)--FSO   & Dual-hop & -- & -- & SER \\ 
\cite{odeyemi2022performance} & RF (RIS)--RF/FSO  & Hybrid dual-hop & Hard switching & -- & OP, ABER \\ 

Proposed model & FSO/THz--RF (RIS)  & Hybrid dual-hop & Hard and soft switching & \checkmark & OP, ABER \\ 
\hline
\end{tabular}
``\end{table*}

\begin{table*}[!h]
\centering
\caption{Comparative Analysis of the Proposed Model and Existing Approaches}
\begin{tabular}{|>{\centering\arraybackslash}p{0.03\linewidth}|>{\centering\arraybackslash}p{0.10\linewidth}|>{\centering\arraybackslash}p{0.27\linewidth}|>{\centering\arraybackslash}p{0.35\linewidth}|}
\hline
\textbf{Ref.} & \textbf{System model} & 
{\textbf{Main contributions and limitations}} & 
{\textbf{Differences with our work}} \\ \hline
\cite{r9} &  RIS-assisted FSO-RF system & \begin{itemize}
    \item  The authors considered a UAV as the source in the FSO link and deployed a RIS in the RF link within their proposed framework. 
\item Since the FSO link is highly directional in nature, it is highly susceptible to blockage caused by obstacles. Consequently, the proposed model faces considerable challenges in practical deployment scenarios. 
\end{itemize} 
& \begin{itemize}
    \item In our proposed model, we consider a parallel FSO/THz link, where identical information is transmitted simultaneously through both links. This approach enhances communication reliability and addresses the limitations of existing works.  Furthermore, we employ a dual-threshold-based soft switching technique. This method mitigates the frequent switching problem observed in previous studies, thereby significantly improving overall system performance. 
 
\end{itemize}\multirow{3}{*}{} \\ \cline{1-3}
\cite{sharma2022performance} &  RIS-assisted hybrid FSO/RF system  & \begin{itemize}
  \item                  The authors proposed a RIS-based hybrid communication framework aimed at improving the reliability of the network. 
\item They employed a maximal ratio combining (MRC)–based switching technique that considered only a single threshold for operation. 
\item  Their proposed model is applicable only for short-distance communication, limiting its practical deployment in larger-scale networks. 

\end{itemize} &       \begin{itemize}
    \item  We consider a dual-hop network instead of a single-hop configuration, which enables reliable long-distance information transmission. Therefore, our proposed model better reflects practical deployment scenarios and addresses the limitations of existing approaches. 
\end{itemize}           \\ \cline{1-3}
\cite{odeyemi2022performance} &  RIS-empowered dual-hop hybrid RF-FSO/RF system & \begin{itemize}
\item The authors analyzed the performance of a RIS-assisted mixed model, where the RIS was installed on the rooftop of a building. 
\item They adopted a selection combining (SC)- switching technique that operated with a single threshold value.
\item Since the RIS is installed on the rooftop of a building, it cannot change the position to adapt to varying link conditions.  In addition, the use of a single-threshold switching causes frequent switching between links, which considerably deteriorates the performance. 

\end{itemize} &        \begin{itemize}
    \item Our proposed model considers a UAV-assisted RIS that can change its optimal placement to maintain reliable LoS links between the transmitter and receiver. This capability makes the system particularly suitable for urban, rural and disaster recovery environments. 
\end{itemize}           \\ \hline
\end{tabular}
\end{table*}
In this paper, we introduce a novel RIS-aided aerial communication model for a dual-hop hybrid FSO/THz relaying system, which is crucial for NTN-based $6$G network. In this model, the FSO link experiences Málaga turbulence, the THz link follows an $\alpha$-$\mu$ distribution, and the RF link undergoes Rician fading. An intermediate access point receives both FSO and THz signals simultaneously from the source, performs switching, and transmits it to the mobile user via a RIS-mounted aerial vehicle. However, this research provides following key contributions:
\begin{enumerate}
\item \textbf{Introducing a Reliable Hybrid Communication Framework with Soft Switching:} Although several studies \cite{n2, n3, n23, n4, n5} have proposed hybrid communication models, they have focused on hard switching techniques only. In contrast, our work assumes soft switching at the receiver of the hybrid network to reduce power wastage. Meanwhile, some studies \cite{ris,mahmoud2021intelligent} have incorporated aerial RIS into their models; however, they do not consider hybrid architectures, which could affect the reliability of communication. Therefore, our proposed system is fundamentally different in terms of configuration and offers a more reliable and efficient communication framework.

\item \textbf{Derivation of Novel Closed-Form Expressions for Performance Metrics:} We utilize the probability density function (PDF) and cumulative distribution function (CDF) of each link (FSO, THz, and RF) to derive the dual-hop CDF expressions for both hard and soft switching schemes. These analytical CDF expressions are then used to obtain CFEs for key performance metrics, such as OP and ABER, under both hard and soft switching mechanisms. The derived expressions are novel, as they correspond to a unique system model featuring RIS-based aerial setup implementation for a dual-hop hybrid network.
\item \textbf{Comprehensive Performance Evaluation under Practical Conditions:} Using the derived analytical expressions, we analyze the impact of key factors, including point errors, atmospheric turbulence, the number of RIS reflecting elements, and the position and altitude of the UAV. Furthermore, we conduct a comparative evaluation of different detection techniques, such as HD and IM/DD, as well as switching methods, including soft and hard switching.
\item \textbf{Asymptotic and Simulation-Based Validation:} To further strengthen the analytical findings, we perform asymptotic analysis in the high-SNR regime and also determine the diversity order of the proposed system. Finally, extensive Monte Carlo (MC) simulations are conducted to validate the accuracy of the derived analytical expressions.
\end{enumerate}
The structure of the paper is organized as follows: Section \ref{sec2} introduces the system model and presents the problem formulation. Section \ref{sec3} derives closed-form expressions for key performance metrics, including outage probability (OP) and average bit error rate (ABER), along with their asymptotic analyses. Section \ref{sec4} provides numerical and simulation results, offering insights relevant to practical implementation. Finally, Section \ref{sec6} concludes the paper by summarizing the main contributions.
\section{System Model and Problem Formulation}
\label{sec2}
\begin{figure}[!h]
\vspace{0mm}
\centerline{\includegraphics[width=0.5\textwidth,angle=0]{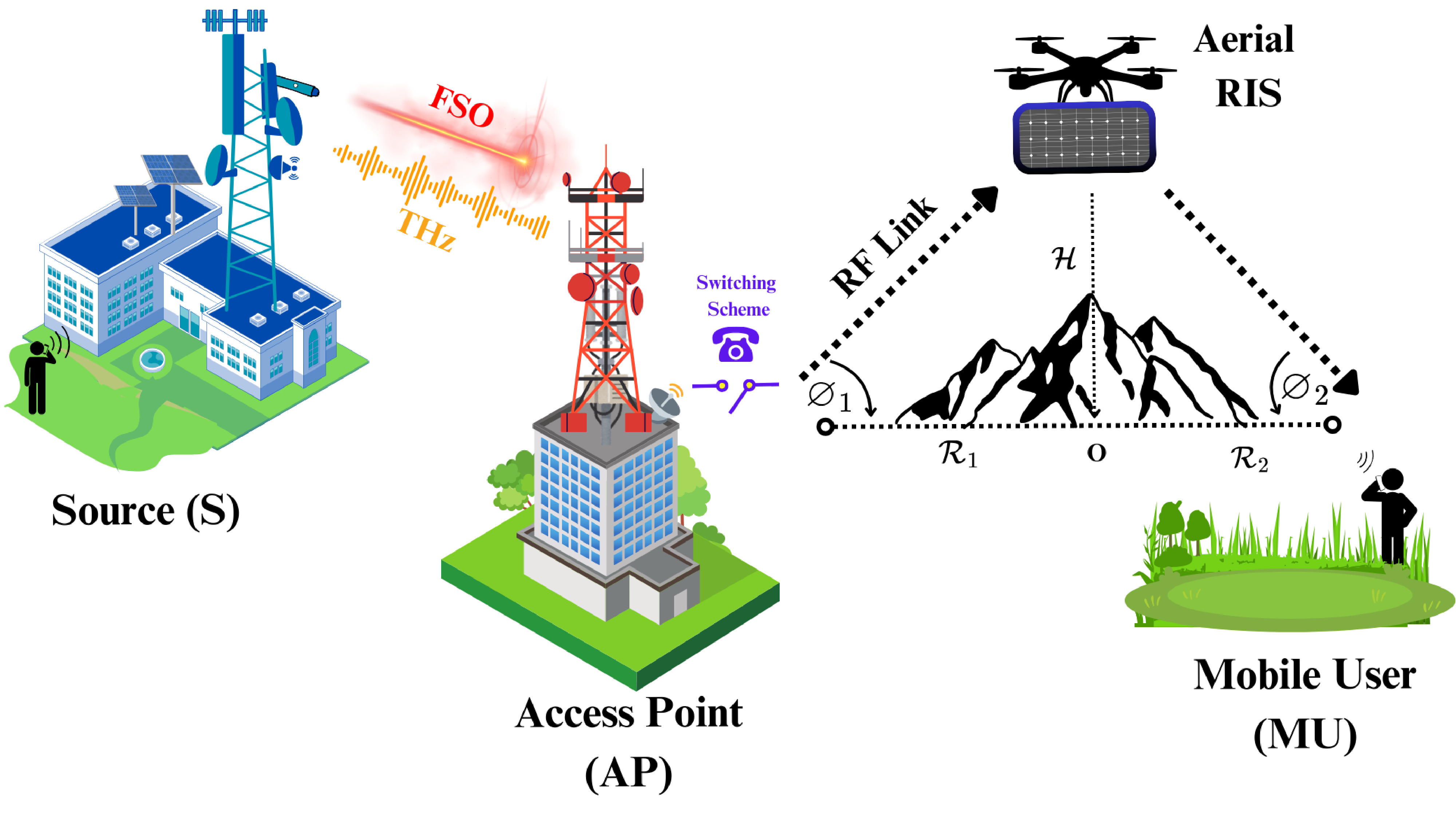}}
    \vspace{0mm}
    \caption{Proposed system model incorporating a source, a switching center, an aerial RIS, and a receiver.}
    \label{fig 1}
\end{figure}
Fig. \ref{fig 1} demonstrates a mixed dual-hop network providing high data rates to a terrestrial user through RIS, featuring a source, $S$ (e.g., ground control station, smartphone), an access point, $AP$ (e.g., cell phone tower), a RIS mounted on an UAV (e.g., drone) and a mobile user, $MU$. The first hop consists of a mixed backhaul network where $S$ transmits information simultaneously to $AP$ via two different links (i.e. the FSO link and the THz link). Here, the $S-AP$ link acts as the foundation of the network, enabling seamless communication between two points. At $AP$, a switching mechanism compares the signal strengths received from both links to determine the optimal transmission link. Given its ability to provide high data rate secure communication, the FSO link is prioritized over the THz link by default. However, if the performance of the FSO link deteriorates, the THz link is utilized through $AP$. A decode and forward (DF) relay system at $AP$ converts the selected signal to RF signal. However, considering a significant distance ($\mathcal{R}_1-O-\mathcal{R}_2$) between $AP$ and $MU$, along with obstacles such as tall mountains and buildings, signal degradation occurs. To address these challenges and mitigate path loss, the RF signal is directed to the aerial RIS, which is positioned at a height $\mathcal{H}$ above the ground, for signal reflection towards $MU$. The aerial RIS serves as an intermediary, reflecting signals from $AP$ to $MU$ while dynamically adjusting its elevation to overcome obstacles and minimize signal attenuation. The FSO link is assumed to experience Málaga turbulence, the THz link undergoes $\alpha-\mu$ distribution, and the RF link follows a Rician distribution. 

\textbf{Remark 1.} \textit{Such hybrid models demonstrate their adaptability and benefits across various applications, including scenarios like natural disasters, densely populated urban areas, addressing handover issues for IoT devices, minimizing dropped calls in metropolitan regions, military operations, and countering potential eavesdropping attempts \cite{ibrahim}.}

\subsection{SNRs of Individual Link}
\noindent In the initial stage of transmission, both the FSO and THz links transmit identical information from $S$ to $AP$ links simultaneously. However, the received signal expression at the $AP$ for the FSO link can be represented as
\begin{align}
     y_o=\mathcal{I}  \eta _e x +w_f,
\end{align}
where \(\eta _e\) is the photoelectric conversion ratio, $x$ is the transmitted signal, \(\mathcal{I}\) is the receiver irradiance given as \(\mathcal{I}=\mathcal{I}_a \mathcal{I}_l \mathcal{I}_p\), \(\mathcal{I}_l\) represents the exponential route loss component, \(\mathcal{I}_a\) represents the channel random irradiance due to air turbulence, \(\mathcal{I}_p\) represents the pointing error impairment irradiance, and \(w_f\) represents the additive white gaussian noise (AWGN) process with a power spectral density of \(N_0\) Watts/Hz.
Finally, the instantaneous received SNR for the FSO link can be written as $\gamma_o=\frac{\left(\mathcal{I} \eta _e\right)^{s}}{N_0}$, where \(s\) denotes the receiver detection technique and the electrical SNR is defined as \(\mu_{o_{s}}=\frac{\eta _e^{s} \mathbb{E}^{s}(\mathcal{I})}{N_0}\), \(\mathbb{E}[\cdot]\) represents the expectation operator \cite{r15}.
Similarly, the received signal expression at the $AP$ for the THz link can be written as
\begin{align}
    y_t=x \sqrt{P_t} \mathit{h}_t\left(\sum _{i=1}^{N_t} h_{f_i} h _{p_i}\right)+w_t,
\end{align}
where \(P_t\) represents the source power, \(\mathit{h}_t\) denotes route loss, \(N_t\) is the number of antennas at the THz receiver,  \(h_{p}\) denotes the antenna misalignment, \(h_{f}\) indicates the channel fading at the $i$-th antenna, and \(w_t\) represents the AWGN with a mean of 0 and a variance \(\varpi _t^2\). Using the maximal ratio technique, the instantaneous SNR at the THz receiver can be expressed as $\gamma _t=\overline{\gamma }_t\begin{Vmatrix} h_t \end{Vmatrix}^2$, where \(\overline{\gamma }_t=\frac{P_t \mathit{h}_t^2}{\varpi _t^2}\) is the average received SNR of the THz link and \(\begin{Vmatrix} h_t \end{Vmatrix}^2= h_p^2 \sum _{i=1}^{N_t} \left[h_{f_i}\right]^2\). Now, the second hop includes a UAV equipped system with a RIS composed of \(N_r\) reflecting elements. In this work, we consider the far-field case where \(\mathcal{L}_1\) is the distance from $AP$ to UAV and  \(\mathcal{L}_2\) is the distance from UAV to $MU$s. The distance can be demonstrated as $\mathcal{L}_k=\sqrt{\mathcal{R}^2_k+\mathcal{H}^2}$, where \(k\in \{1,2\}\), \(\mathcal{H}\) is the altitude of the UAV, \(\mathcal{R}_1\) and \(\mathcal{R}_2\) are the projection distances between $AP$ to UAV and UAV to $MU$. Now, the received signal at $MU$s can be formulated as
\begin{align}
\label{r snr 1}
    y_r= \sqrt{\frac{P_s}{\mathcal{L}_1^{\mathit{i}_1} \mathcal{L}_2^{\mathit{i}_2}}}\left(\sum _{k=1}^{N_r} \delta _k \zeta _k e^{\text{j$\Theta $}_k} e^{-\text{j$\Psi $}_k} e^{-\text{j$\vartheta $}_k}\right)x+w_r,
\end{align}
\noindent where \(P_s\) denotes average energy per symbol, (\(\mathit{i}_1,\mathit{i}_2\)) are the corresponding path loss components, \(w_r\sim \mathcal{N}\left(0,N_d\right)\) is the AWGN, \(\text{$\Theta $}_k\) is the phase shift applied by k-th reflecting element of the RIS, \(\vartheta _k\) and \(\Psi _k\) are the phases of the channel coefficients, respectively, \(\delta _k\) and \(\zeta _k\) denote channel amplitudes of Rician distribution, respectively. Therefore, the instantaneous SNR at $MU$ can be written as
\begin{align}
    \gamma _r=\frac{P_s\left(\sum _{k=1}^{N_r} \delta _k \zeta _k e^{j \left(\Theta _k-\Psi _k-\vartheta _k\right)}\right)^2}{N_d \mathcal{L}_1^{\mathit{i}_1} \mathcal{L}_2^{\mathit{i}_2}}.
\end{align}
In UAV-assisted systems, the aerial RIS can adjust its position to maintain LOS conditions and ensure that the required phase alignment condition holds dynamically, even under channel fluctuations or environmental turbulence \cite{10697101}. Therefore, to maximize \(\gamma _r\), the phase shift generated by the RIS is set to \(\Theta_k=\Psi _k+\vartheta _k\). Accordingly, the maximum SNR can be expressed as $\gamma _r=\frac{\overline{\gamma }_r(\mathcal{Y})^2}{\mathcal{L}_1^{\mathit{i}_1} \mathcal{L}_2^{\mathit{i}_2}}$, where \(\overline{\gamma }_r\) is the average SNR for the RF link and is denoted by \(\overline{\gamma }_r=\frac{P_s}{N_d}\) and \(\mathcal{Y}=\sum _{k=1}^{N_r} \delta _k \zeta _k\).

\subsection{PDF and CDF of FSO Link}
\footnote{Málaga turbulence model is a well-known generalized model that incorporates some classical turbulence models as its special instances \cite{ibrahim}.}We assume that the FSO link is subjected to the Málaga distribution with pointing error impairments. Therefore, the PDF and CDF of \(\gamma _o\) can be expressed, respectively, as \cite[Eqs. (12) and (13)]{ibrahim}
\begin{align}
\label{pdf m}
f_{\gamma _o} (\gamma)&=\frac{\Omega _o \epsilon _o^2}{2^{s} \gamma}\sum _{z=1}^{\beta _o} v_z G_{1,3}^{3,0}\left[\Lambda_0 \left(\frac{\gamma}{\mu _{o_s}}\right)^{\frac{1}{s}}\Bigg|\begin{array}{c}\epsilon _o^2+1 \\\epsilon _o^2,\alpha _o,z \\\end{array}\right],\\
\label{malaga}
F_{\gamma _o}( \gamma )&=\frac{\Omega _o \epsilon _o^2}{2^{2 s-1} \pi ^{s-1}}\sum _{z=1}^{\beta _o} w_{z} G_{ s+1, 3 s+1}^{ 3 s ,1 } \left [ \frac{\mathcal{Z} \gamma}{\mu _{o_s}}  \Bigg| 
\begin{array}{cc}
 1,\mathcal{T}_1 \\
 \mathcal{T}_2,0 \\
\end{array}
 \right],
\end{align}
where
$\Omega _o=\frac{2 \alpha _o^{\frac{\alpha _o}{2}}}{\Gamma \left(\alpha _o\right) \mathcal{J}_p^{\frac{\alpha _o}{2}+1}}\left(\frac{\beta _o \mathcal{J}_p}{\beta _o \mathcal{J}_p+\text{\' o}}\right)^{\frac{\alpha _o}{2}+\beta _o}$, $\Lambda_0=\frac{\alpha _o \beta _o \epsilon _o^2 \left(\mathcal{J}_p+\text{\' o}\right)}{\left(\epsilon _o^2+1\right) \left(\beta _o \mathcal{J}_p+\text{\' o}\right)}$, $v_z=u_z \left(\frac{\alpha _o \beta _o}{\beta _o \mathcal{J}_p+\text{\' o}}\right)^{-\frac{1}{2} \left(\alpha _o+z\right)}$, $u_z=\frac{1}{(z-1)!} \left(
\begin{array}{c}
\beta _o-1 \\
z-1 \\
\end{array}
\right) \left(\frac{\alpha _o}{\beta _o}\right)^{\frac{z}{2}}\left(\frac{\text{\' o}}{\mathcal{J}_p}\right)^{z-1} \left(\beta _o \mathcal{J}_p+\text{\' o}\right)^{1-\frac{z}{2}}$,  \(w_{z}=v_z s^{\alpha _o+z-1}\), \(\mathcal{Z} =\frac{\Lambda_0^{s}}{s^{2 s}}\), \(\mathcal{T}_1=\left\{\frac{\epsilon _o^2+1}{s},\text{...}.,\frac{\epsilon _o^2+s}{s}\right\}\), 
\(\mathcal{T}_2=\left\{\frac{\epsilon _o^2}{s},\text{...},\frac{\epsilon _o^2+s-1}{s},\frac{\alpha _o}{s},\text{...}.,\frac{\alpha _o+s-1}{s},\frac{z}{s},\text{...}.,\frac{z+s-1}{s}\right\}\), (\(\alpha _o\), \(\beta _o\)) denote the turbulence parameter, \(\epsilon _o\) indicates the pointing error impairments, and
\(s\in \{1,2\}\) represents the detection technique. Here, \(s\) = $1$ denotes the HD technique and \(s\) = $2$ represents the IM/DD technique, \(\mu _{o_s}\) defines the electrical SNR related to the average SNR of optical link ($\overline{\gamma }_{o})$, \(\mu _{o_1}\) = \(\overline{\gamma }_{o}\) for the HD method, \(\mu _{o_2} =\frac{\overline{\gamma }_{o}\alpha _o \epsilon _o^2 \left(\epsilon _o^2+2\right) (\text{\' o} \mathcal{J}_p)}{(\epsilon _o^2+1)^2 [\text{\' o}^2(\alpha _o+1)  (\frac{1}{\beta _o}+1)+2 \mathcal{J}_p (\mathcal{J}_p+2 \text{\' o})]}\) for the IM/DD technique. The average power associated with scattering component through interaction with  off-axis eddies, is presented by the symbol \(\mathcal{J}_p=2 \mathbf{c}_0 \left(1-\rho _0\right)\), the average power of all scattered components is denoted by the term \(2 \mathbf{c}_0\), the parameter \(0\leq \rho _0\leq 1\) represents the fraction of scattered power that is coherently coupled to the LOS component, the average power of the combined coherent contributions is presented by \(\text{\' o}=2 \cos  \left(\mho _A-\mho _B\right) \sqrt{2 \mathbf{c}_0 \text{\" o} \rho _0}+2 \mathbf{c}_0 \rho _0+\text{\" o}\), \(\text{\" o}\) indicates the average power of the LOS components, and the deterministic phases corresponding to the LOS components are \(\mho _A\) and \(\mho _B\), respectively.

\subsection{PDF and CDF of THz Link}
\footnote{
The $\alpha-\mu$ distribution serves as a highly flexible and accurate model for characterizing wireless fading channels. Its versatility allows it to represent a wide range of multipath environments effectively \cite{roisul}.}We assume that the THz link undergoes $\alpha-\mu$ fading distributions due to its generic advantages and close to the practical environment. Hence, the PDF and CDF of \(\gamma _t\) are expressed, respectively as \cite[Eqs. (12) and (20)]{rakib}
\begin{align}
\label{pdf t}
f_{\gamma _t}(\gamma ) &= \frac{\chi_1}{2  \left(N_t \overline{\gamma }_t\right) ^{\mathit{g}_t}}\gamma^{\mathit{g}_t-1}\Gamma \left[\chi_2,\chi_3 \left(\frac{\gamma }{  N_t \overline{\gamma }_t}\right)^{\frac{\alpha_t}{2}}\right],\\
\label{thz}
F_{\gamma _t}(\gamma) &= \frac{\chi_1 \gamma^{\mathit{g}_t}}{\alpha_t (N_t \overline{\gamma}_t)^{\mathit{g}_t}}G_{2,3}^{2,1}\left[\chi_3 \left(\frac{\gamma}{N_t\overline{\gamma}_t}\right)^{\frac{\alpha_t}{2}}\Bigg|\begin{array}{c}1-\frac{2 \mathit{g}_t}{\alpha_t},1 \\0,\chi_2,-\frac{2 \mathit{g}_t}{\alpha_t}\end{array}\right],
\end{align}
where  $\chi_1=\frac{2 \mathit{g}_t \left(\mu_t  N_t\right)^{\frac{2 \mathit{g}_t}{\alpha_t }}}{\Omega _t^{2 \mathit{g}_t} m^{2 \mathit{g}_t} \Gamma \left(N_t \mu_t \right)}$ , $\chi_2=\frac{\alpha_t  \mu_t  N_t-2 \mathit{g}_t}{\alpha_t }$, $\chi_3=\frac{\mu_t  N_t}{\Omega _t^{\alpha_t } m^{\alpha_t }}$, and   $\mathit{g}_t = \frac{\epsilon_t^2}{2}$. Here $m$ is the \(\alpha\)-root mean value of the fading channel, $\Omega _t$ is the fraction of collected power, \(\alpha_t\) and \(\mu_t\) are the fading parameters, \(\epsilon_t\) is the pointing error parameter, and \(\Gamma (a,z)=\int _z^{\infty } t^{a-1} e^{-t}dt\) represents the upper incomplete gamma function \cite{roisul}.

\subsection{PDF and CDF of RF Link}
\footnote{Rician distribution accurately models realistic propagation environments characterized by a dominant line-of-sight (LoS) path coexisting with multiple scattered components. It is widely used in modeling UAV and RIS-assisted systems and is particularly important for NTNs, where LoS conditions are prevalent \cite{ris}.} We assume that all the RF links undergo Rician distributions. Therefore, the PDF and CDF of \(\gamma _r\) are expressed, respectively, as \cite [Eq. (9) and (10)]{ris}
\begin{align}
\label{pdf r}
f_{\gamma _r}(\gamma)&=\frac{\gamma ^{\frac{1}{2} \left(v_r-1\right)} e^{-\sqrt{\psi _r \gamma}}}{2 \Gamma  \left(v_r+1\right) \psi _r^{-\frac{1}{2} \left(v_r-1\right)}},\\
\label{ris}
F_{\gamma _r}(\gamma  )&=\frac{\Gamma \left(v_r+1,\sqrt{\psi _r \gamma}\right)}{\Gamma\left(v_r+1\right)},
\end{align}
where $\psi _r=\frac{P_{L_r}}{\overline{\lambda} ^2_r \overline{\gamma }_r}$, \(\overline{\lambda} _r\) and \( v_r\) are related to the mean and variance of the Rician random variable (\( \mathcal{Y}\)), $\overline{\lambda} _r=\frac{V(\mathcal{Y})}{\mathbb{E}(\mathcal{Y})}$, and $v_r=\frac{\mathbb{E}(\mathcal{Y})^2}{V(\mathcal{Y})}-1$. The mean and variance are expressed as $\mathbb{E}(\mathcal{Y})=N_r\prod _{k=1}^2 \mathbb{E}_k(\mathcal{Y})$ and \(V(\mathcal{Y})=N_r \left(\prod _{k=1}^2 V_k(\mathcal{Y})-\mathbb{E}(\mathcal{Y})^2\right)\), respectively. Here, $   \mathbb{E}_k(\mathcal{Y})=\frac{3 \Gamma  \left(_1 F_1\left(\frac{3}{2};1;J_k\right)\right)}{2 \left(\sqrt{J_k+1} e^{J_k}\right)}$ and $V_k(\mathcal{Y})=\frac{_1 F_1\left(2;1;J_k\right)}{\sqrt{J_k+1} e^{J_k}}-\mathbb{E}_k(\mathcal{Y})^2$, \(J_k=\mathit{b}_1 e^{\mathit{b}_2 \varnothing _k}\) is the Rician Factor, \(\mathit{b}_1=J(0)\), \(\mathit{b}_2=\frac{2}{\pi} \log \left[\frac{J\left(\frac{\pi}{2}\right)}{J(0)}\right]\), the maximum and minimum values of Rice factor are presented by \(J\left(\frac{\pi}{2}\right)\) and \(J(0)\), respectively \cite{shimamoto2006channel}, and 
\(_1 F_1(\cdot)\) is the Kummer confluent hypergeometric function as defined in \cite{edition2007table}.
Therefore,  the path loss of \(\gamma _r\) can be expressed as $P_{L_r}=\mathcal{L}_1^{\mathit{i}_1} \mathcal{L}_2^{\mathit{i}_2}$, where \(\mathit{i}_k=\mathit{b}_4+\mathit{b}_3 \text{ } \mathcal{U} (\varnothing _k)\) denotes the path loss exponent for \(k\in \{1,2\}\), \(\mathcal{U} (\varnothing _k)=\frac{1}{\mathit{b}_5 e^{-\mathit{b}_6 \left(\varnothing _k-\mathit{b}_5\right)}+1}\) indicates the probability of LOS. Here, \(\mathit{b}_3=\delta _{\frac{\pi }{2}}-\delta _0\), \(\mathit{b}_4=\delta _0\), and (\(\mathit{b}_5,\mathit{b}_6\)) represent the constants related to environment and transmission frequency \cite{lei2019secure}. The values of (\(\mathit{b}_5,\mathit{b}_6\)) are found to be ($4.88, 0.43$) in suburban areas, ($9.61, 0.16$) in urban areas, ($12.08, 0.11$) in dense urban areas, and ($27.23, 0.08$) in high-rise urban areas \cite{lei2019secure}. Now, the term \(\varnothing _k\) represents the angles between the UAV and the $AP$, and between the UAV and the $MU$, respectively, which is denoted as \(\varnothing _k=\tan ^{-1}\left(\mathcal{R}_k,\mathcal{H}\right)\).
\setcounter{eqnback}{\value{equation}}
\setcounter{equation}{18}
\begin{figure*}
\begin{align}
\label{eq 36}
P_{out}^H(\gamma_{th},\gamma_{th}^r)&=
\sum _{z=1}^{\beta _o} \mathcal{V}_{1}(\gamma _{\text{th}})^{\mathit{g}_t}w_{z} G_{ s+1, 3 s+1}^{ 3 s ,1 } \left [ \frac{\mathcal{Z}\gamma _{\text{th}} }{\mu _{o_s}}  \Bigg| 
\begin{array}{cc}
 1 , \mathcal{T}_1 \\
 \mathcal{T}_2 , 0 \\
\end{array}
 \right]\nonumber
 G_{2,3}^{2,1}\left[\chi_3 \left(\frac{\gamma _{\text{th}} }{\overline{\gamma }_t N_t}\right)^{\frac{\alpha_t }{2}}\Bigg| 
\begin{array}{c}
 1-\frac{2 \mathit{g}_t}{\alpha_t },1 \\
 0,\chi_2,-\frac{2 \mathit{g}_t}{\alpha_t } \\
\end{array}
\right] 
+\frac{\Gamma \left(v_r+1,\sqrt{\psi _r \gamma _{\text{th}}^r}\right)}{\Gamma\left(v_r+1\right)}
\nonumber
\\
&\times 
\Bigg\{1-
\sum _{z=1}^{\beta _o} \mathcal{V}_1(\gamma _{\text{th}})^{\mathit{g}_t}w_{z} G_{ s+1, 3 s+1}^{ 3 s ,1 } \left [ \frac{\mathcal{Z}\gamma _{\text{th}} }{\mu _{o_s}}  \Bigg| 
\begin{array}{cc}
 1 , \mathcal{T}_1 \\
 \mathcal{T}_2 , 0 \\
\end{array}
 \right] G_{2,3}^{2,1}\left[\chi_3 \left(\frac{\gamma _{\text{th}} }{\overline{\gamma }_t N_t}\right)^{\frac{\alpha_t }{2}}\Bigg| 
\begin{array}{c}
 1-\frac{2 \mathit{g}_t}{\alpha_t },1 \\
 0,\chi_2,-\frac{2 \mathit{g}_t}{\alpha_t } \\
\end{array}
\right]\Bigg\}.
\end{align}
\rule{\textwidth}{0.5pt}
\end{figure*}
\subsection{CDF of hybrid FSO/THz link for Hard Switching}
In our proposed model, the FSO link is given higher priority than the THz link due to its superior data rate capabilities
Therefore, the hard switching in hybrid FSO/THz link is performed on the basis of the following conditions.
\setcounter{eqnback}{\value{equation}}
\setcounter{equation}{10}
\begin{align}
\gamma _h= \left\{\begin{matrix}
\gamma _o, & if \textrm{ }  \gamma _o\geq \gamma_{th} \\ 
\gamma _t, &if \textrm{ } \gamma _o<\gamma_{th} ,\gamma _t\geq \gamma_{th} \\ 
 0,& if \textrm{ }  \gamma _o<\gamma_{th},\gamma _t<\gamma_{th}.
\end{matrix}\right.
\end{align}
Where $\gamma_{th}$ is the predefined threshold value and $\gamma_h$ denotes the instantaneous SNR of the FSO/THz link. 

\textbf{Remark 2.} \textit{Hard switching approach relies on a single threshold criterion to select between the FSO and THz links. The FSO link is activated when its SNR exceeds a predefined threshold. If this condition is not met, the receiver evaluates the THz link and activates it, provided its SNR meets or exceeds the required threshold.}

Therefore, the CDF of $\gamma_h$ can be written as \cite[Eq. (20)]{ibrahim}
\begin{align}
\label{hybrid}
   F_{\gamma _h}(\gamma_{th}) &=F_{\gamma _o}(\gamma_{th}) \times F_{\gamma _t}(\gamma_{th}).
\end{align}
Substituting \eqref{malaga} and \eqref{thz} into \eqref{hybrid}, $F_{\gamma _h}(\gamma_{th})$ is obtained as
\begin{align}
\label{first hop}
 F_{\gamma _h}(\gamma_{th}) &=\frac{\chi_1\Omega _o \epsilon _o^2}{\alpha_t  \left(N_t \overline{\gamma }_t\right)^{\mathit{g}_t}2^{2 s-1} \pi ^{s-1}}
 \nonumber 
 \\
 &\times\sum _{z=1}^{\beta _o} w_{z} (\gamma_{th}) ^{\mathit{g}_t}G_{ s+1, 3 s+1}^{ 3 s ,1 } \left [ \frac{\mathcal{Z} \gamma_{th} }{\mu _{o_s}}  \Bigg| 
\begin{array}{cc}
 1 , \mathcal{T}_1 \\
 \mathcal{T}_2, 0 \\
\end{array}
 \right]\nonumber\\
 &\times G_{2,3}^{2,1}\left[\chi_3 \left(\frac{\gamma_{th}}{\overline{\gamma }_t N_t}\right)^{\frac{\alpha_t}{2}}\Bigg| 
\begin{array}{c}
 1-\frac{2 \mathit{g}_t}{\alpha_t },1 \\
 0,\chi_2,-\frac{2 \mathit{g}_t}{\alpha_t } \\
\end{array}
\right].
\end{align}

\subsection{CDF of hybrid FSO/THz link for Soft Switching}
In contrast to the hard switching approach, which relies on a single threshold, this method utilizes a dual-threshold mechanism for the FSO link and a single threshold for the THz link. As a result, the FSO link remains active under a broader range of conditions. Thus, the soft switching in hybrid FSO/THz link is executed based on the following criteria.
\begin{align}
    \gamma _s=\left\{\begin{matrix}
    \gamma _o, &if  \textrm{ }  \gamma _o\geq \gamma_{th}^u\\ 
\gamma _o, &if\textrm{ } \gamma_{th}^l\leq \gamma _o\leq \gamma_{th}^u,\gamma _o\geq \gamma_{th}^u \\ 
\gamma _t, &if\textrm{ } \gamma_{th}^l\leq \gamma _o\leq \gamma_{th}^u\text{},\gamma _o<\text{}\gamma_{th} ^{l\text{}\text{}}\text{},\gamma _t\geq \text{}\gamma_{th}^t \\ 
\gamma _t, &if\textrm{ }\gamma _o<\text{}\gamma_{th}^{l},\gamma _t\geq \text{}\gamma_{th}^t 
\\
0, &if\textrm{ } \gamma_{th}^l\leq \gamma _o\leq \gamma_{th}^u,\text{}\gamma _o<\text{}\gamma_{th} ^{l\text{}\text{}},\gamma _t<\text{}\gamma_{th}^t\\
0, & if\textrm{ } \gamma _o<\gamma_{th}^{l},\gamma _t<\gamma_{th}^t.
\end{matrix}\right.
\end{align}
Where (\(\gamma_{th}^u\), \(\gamma_{th}^l\)) denotes the upper and lower thresholds of FSO link, \(\gamma_{th}^t\) denotes the threshold of THz link, and $\gamma_s$ denotes the instantaneous SNR of the hybrid link. 

\textbf{Remark 3.} \textit{To address the issue of unnecessary and frequent switching, a soft switching strategy incorporating adaptive thresholding is introduced. Soft switching technique offers configurational flexibility as it can assign two thresholds to the THz link instead of the FSO link when channel conditions require. This adaptability enables the switching mechanism to respond more effectively to dynamic and unpredictable variations in the communication environment.}

Therefore, the CDF of $\gamma_s$ can be obtained as
\begin{align}
\label{soft hybrid}
   F_{\gamma _s}\left(\gamma_{th}^u,\gamma_{th}^l, \gamma_{th}^t\right)&=F_{\gamma _o}\left(\gamma_{th}^u,\gamma_{th}^l\right)\text{ }\times \text{ }F_{\gamma _t}(\gamma_{th}^t).
\end{align}
Here, \(F_{\gamma _o}\left(\gamma_{th}^u,\gamma_{th}^l\right)\) can be written as
\begin{align}   
\label{soft m}
    F_{\gamma _o}\left(\gamma_{th}^u,\gamma_{th}^l\right)&=F_{\gamma _o}(\gamma_{th}^l)+\frac{F_{\gamma _o} (\gamma_{th}^l) \left[F_{\gamma _o}(\gamma_{th}^u)-F_{\gamma _o}(\gamma_{th}^l)\right]}{F_{\gamma _o}(\gamma_{th}^l)-F_{\gamma _o} (\gamma_{th}^u)+1},
\end{align}
where \(F_{\gamma _o} (\gamma_{th}^l)\) and \(F_{\gamma _o} (\gamma_{th}^u)\) are obtained from \eqref{malaga} and \(F_{\gamma _t}(\gamma_{th}^t)\) is obtained from \eqref{thz} by applying appropriate thresholds. The term \(F_{\gamma _o} (\gamma_{th}^l)\) represents the probability that \(\gamma _o\) is less than lower threshold \(\gamma_{th} ^{l}\). The difference \([F_{\gamma _o} (\gamma_{th}^u)-F_{\gamma _o} (\gamma_{th}^l)]\) corresponds to the probability that \(\gamma _o\) lies between lower annd upper thresholds, i.e., \(\gamma_{th}^l\leq \gamma _o\leq \gamma_{th}^u\). Similarly, \([1-F_{\gamma _o} (\gamma_{th}^u)]\) denotes the probability that \(\gamma _o\) is greater than or equal to the upper threshold \(\gamma_{th} ^{u}\). Provided that \(\gamma _o < \gamma_{th} ^{l}\), the expression \(\frac{ F_{\gamma _o} (\gamma_{th}^u)-F_{\gamma _o}( \gamma_{th}^l)}{F_{\gamma _o} (\gamma_{th}^l)-F_{\gamma _o} (\gamma_{th}^u)+1}\) represents the probability that \(\gamma _o\) lies within the interval \([\gamma_{th}^l,\gamma_{th}^u]\).
\section{Performance Analysis}
\label{sec3}
\subsection{OP Analysis}
OP in wireless communication systems refers to the risk of a communication outage or failure if the received signal quality falls below a preset threshold (\(\gamma _{\text{th}}\)). It is a significant metric for evaluating
and creating dependable wireless networks. However, the OP is mathematically defined as \cite[Eq. (39)]{rakib}
\begin{align}
    P_{out}&=P_r \left\{\gamma <\gamma _{\text{th}}\right\}=F_{\text{eq}} (\gamma _{\text{th}}).
\end{align}
\subsubsection{Hard Switching}
The FSO/THz hybrid link experiences outage when \(\gamma _o\) and \(\gamma _t\) are less than \(\gamma _{\text{th}}\). Considering dual-hop relaying network, effective communication requires both connections to remain continuous. 
So, the end-to-end outage probability for hard switching technique is given as
\begin{align}
\label{e2e}
    P_{out}^H=1-\left[1- F_{\gamma_h}(\gamma _{\text{th}})\right] \left[1-F_{\gamma _r} (\gamma _{\text{th}}^r)\right],
\end{align}
where \(\gamma _{\text{th}}^r\) is the threshold SNR at the fronthaul network. Substituting \eqref{first hop} and \eqref{ris} into \eqref{e2e}, $ P_{out}^H$ is derived as shown in \eqref{eq 36}, where $\mathcal{V}_1=\frac{\chi_1 \Omega _o \epsilon _o^2}{2^{2 s-1} \pi ^{s-1}\alpha_t  \left(N_t \overline{\gamma }_t\right)^{\mathit{g}_t}}.$
\subsubsection{Soft Switching}
Similar to hard switching approach, the end-to-end outage probability for the soft switching technique is obtained as
\setcounter{eqnback}{\value{equation}}
\setcounter{equation}{19}
\begin{align}
\label{abr}
    P_{out}^S=1-\left[1-  F_{\gamma _s}\left(\gamma_{th}^u,\gamma_{th}^l, \gamma_{th}^t\right)\right] \left[1-F_{\gamma _r} (\gamma _{\text{th}}^r)\right].
\end{align}
Now, putting \eqref{soft hybrid} and \eqref{ris} into \eqref{abr}, $ P_{out}^S$ is obtained as shown in \eqref{eq 37}, where $\mathcal{V}_2=\frac{2^{2 s-1} \pi ^{s-1}\chi_1 (\gamma _{th}^t)^{\mathit{g}_t}}{\Omega _o \epsilon _o^2\alpha_t  \left(N_t \overline{\gamma }_t\right)^{\mathit{g}_t}}$.
\setcounter{eqnback}{\value{equation}}
\setcounter{equation}{20}
\begin{figure*}
\begin{align}
\label{eq 37}
P_{out}^S(\gamma_{th}^u,\gamma_{th}^l,\gamma_{th}^t,\gamma_{th}^r) &= \Bigg\{\frac{\Gamma \left(v_r+1,\sqrt{\psi _r \gamma _{\text{th}}^r}\right)}{\Gamma  \left(v_r+1\right)}+\sum _{z=1}^{\beta _o}\mathcal{V}_2 w_{z}
G_{2,3}^{2,1}\left[\chi_3 \left(\frac{\gamma_{th}^t}{N_t \overline{\gamma }_t}\right)^{\frac{\alpha_t }{2}}\Bigg| 
\begin{array}{c}
 1-\frac{2 \mathit{g}_t}{\alpha_t },1 \\
 0,\chi_2,-\frac{2 \mathit{g}_t}{\alpha_t } \\
\end{array}
\right]
G_{ s+1, 3 s+1}^{ 3 s ,1 } \left [ \frac{\mathcal{Z} \gamma_{th}^l}{\mu _{o_s}}  \Bigg| 
\begin{array}{cc}
 1 , \mathcal{T}_1 \\
 \mathcal{T}_2 , 0 \\
\end{array}
 \right]
 \nonumber
\\
&
+\sum _{z=1}^{\beta _o} w_{z} G_{ s+1, 3 s+1}^{ 3 s ,1 } \left [\frac{\mathcal{Z} \gamma_{th}^u}{\mu _{o_s}}  \Bigg| 
\begin{array}{cc}
 1 , \mathcal{T}_1 \\
 \mathcal{T}_2 , 0 \\
\end{array}
 \right]+G_{ s+1, 3 s+1}^{ 3 s ,1 } \left [ \frac{\mathcal{Z} \gamma_{th}^l}{\mu _{o_s}}  \Bigg| 
\begin{array}{cc}
 1 , \mathcal{T}_1 \\
 \mathcal{T}_2 , 0 \\
\end{array}
 \right]^{-1}\Bigg\} \bigg[1-\frac{\Gamma \left(v_r+1,\sqrt{\psi _r \gamma _{\text{th}}^r}\right)}{\Gamma  \left(v_r+1\right)})\bigg].
\end{align} 
\rule{\textwidth}{0.5pt}
\end{figure*}
\subsection{ABER Analysis}
ABER is a statistical metric that quantifies the probability of errors in bits transferred over a communication channel. It is an important statistic for determining the dependability and performance of digital communication networks. Mathematically, ABER can be expressed as
\begin{align}
\label{ber}
     B_{e}=\sum _{\text{i}=1}^{\mathit{n}} \mathcal{A} \int_{\gamma _{\text{th}}}^{\infty }    \text{erfc} \left(\sqrt{\mathcal{D} _{\text{i}}\gamma }\right) f_{\gamma }(\gamma)\, d\gamma,
\end{align}
where \(f_{\gamma }(\gamma )\) defines the PDF of fading link SNR, \(\text{erfc}\left(t_0\right)\) = \(\frac{2}{\sqrt{\pi }}\int_{t_0}^{\infty } e^{-z^2} \, dz\) denotes the error function, (\(\mathcal{A}\), \(\mathcal{D} _{\text{i}}\)) denotes the modulation techniques, and \(\mathit{n}\) is the summation limit for various modulation scheme. Therefore, the values of \(\mathit{n}\), \(\mathcal{A}\) and \(\mathcal{D} _{\text{i}}\) for the various modulation techniques are given in the following Table III. 
\begin{table}[!htbp]
\renewcommand{\arraystretch}{1.7}
\centering
\caption{Modulation Schemes and Parameters \cite{zedini2020performance}}
\begin{tabular}{| c | c | c | c |}
\hline
Modulation Scheme & \(\mathit{n}\)           & \(\mathcal{A}\)   & \(\mathcal{D} _{\text{i}}\)  \\ \hline
OOK               & $1$           & $\frac{1}{2}$  & $\frac{1}{2}$  \\ \hline
BPSK              & $1$           & $\frac{1}{2}$ & $1$   \\ \hline
M-PSK             & Max \((\frac{M}{4},1)\) &   $\frac{1}{\text{Max}\left(2,\log _2(M)\right)}$   &  $ \sin ^2(\frac{(2 \text{i}-1)\pi }{M})$   \\ \hline
M-QAM             & \(\frac{\sqrt{M}}{2}\)            &  $\frac{2 \left(1-\frac{1}{\sqrt{M}}\right)}{\log _2(M)}$  &  $\frac{3 (2 \text{i}-1)^2}{2 (M-1)}$  \\ \hline
\end{tabular}
\label{mod tab}
\end{table}

\subsubsection{Hard Switching}
The ABER of a dual-hop communication system for hard switching scheme can be written as
\begin{align}
\label{ber1}
   P_\mathbf{e}^H=\text{} P_{\text{hyb}}^H-2P_a P_{\text{hyb}}^H+P_a.
\end{align}
To obtain \(P_\mathbf{e}^H\), the two terms 
\(P_{\text{hyb}}^H\) and \(P_a\) are required, that represent the ABER for the first hop hybrid link and second hop access link, respectively. The derivation of \(P_{\text{hyb}}^H\) and \(P_a\) are given as follows.

\subsection*{Derivation of \(P_{\text{hyb}}^H\):}
\noindent
\(P_{\text{hyb}}^H\) can be written as
\begin{align}
\label{ber hyb h}
    P_{\text{hyb}}^{H}(\gamma _{\text{th}})=\frac{P_t (\gamma _{\text{th}})  F_{\gamma _o}(\gamma _{\text{th}})+P_o(\gamma _{\text{th}})}{1- F_{\gamma _h}(\gamma _{\text{th}})},
\end{align}
where \(P_o(\gamma _{\text{th}})\) and  \(P_t(\gamma _{\text{th}})\) denote the ABER of the FSO and THz links, respectively. \(P_o(\gamma _{\text{th}})\) is obtained by substituting \eqref{pdf m} into \eqref{ber} as
\begin{align}
\label{ber f1}
    P_o (\gamma _{\text{th}})&= \sum _{z=1}^{\beta _o}\sum _{\text{i}=1}^{\mathit{n}} \frac{\mathcal{A} \Omega _o v_z\epsilon _o^2 }{2^{s}}\int _{\gamma _{\text{th}}}^{\infty }\text{erfc}\left(\sqrt{\mathcal{D} _{\text{i}}\gamma }\right)
    \nonumber
    \\
    & \times 
 G_{1,3}^{3,0}\left[\Lambda_0 \left(\frac{\gamma }{\mu _{o_s}}\right)^{\frac{1}{s}}\Bigg| 
\begin{array}{c}
 \epsilon _o^2+1 \\
 \epsilon _o^2,\alpha _o,z \\
\end{array}
\right] \, d\gamma
\nonumber
\\
&=\sum _{z=1}^{\beta _o} \sum _{\text{i}=1}^{\mathit{n}} \frac{\mathcal{A} \Omega _o \epsilon _o^2 v_z}{\sqrt{\pi } 2^{s}}     \left(\mathcal{C}_a+ \mathcal{C}_b\right).
\end{align}
The integral term $\mathcal{C}_a$ is obtained as
\begin{align}
   \mathcal{C}_a&= \int_0^{\infty } \text{erfc}\left(\sqrt{\mathcal{D} _{\text{i}}\gamma }\right) G_{1,3}^{3,0}\left[\frac{\Lambda_0\gamma^{\frac{1}{s}}}{\mu _{o_s}^{\frac{1}{s}}}\Bigg| 
\begin{array}{c}
 \epsilon _o^2+1 \\
 \epsilon _o^2,\alpha _o,z \\
\end{array}
\right] \, d\gamma.
\end{align}
From the conversion of \(\text{erfc}\left(\sqrt{\mathcal{D} _{\text{i}}\gamma }\right)\) into Meijer's $G$ function 
\cite[Eq.(8.4.14.2)]{prudnikov1986integrals} and then utilizing the formula \cite[Eq. (2.24.1.1)]{prudnikov1986integrals}, \( \mathcal{C}_a\) is expressed as
\begin{align}
\label{ra1}
\mathcal{C}_a &=  {(2 \pi )^{1-s} s^{\alpha _o+z-1} }
G_{ s+2, 3 s+1}^{ 3 s ,2 } \left [\frac{\Lambda_0   s^{-2 s} }{\mu _{o_s}\text{}\mathcal{D} _{\text{i}}}   \Bigg| 
\begin{array}{cc}
 1 , \frac{1}{2}, \mathcal{T}_1 \\
 \mathcal{T}_2 , 0 \\
\end{array}
 \right].
\end{align}  
The integral term $\mathcal{C}_b$ is derived as
\begin{align}
     \mathcal{C}_b=\int_0^{\gamma _{\text{th}}} \text{erfc}\left(\sqrt{\mathcal{D} _{\text{i}}\gamma }\right) G_{1,3}^{3,0}\left[\frac{\Lambda_0\gamma^{\frac{1}{s}}}{\mu _{o_s}^{\frac{1}{s}}}\Bigg| 
\begin{array}{c}
 \epsilon _o^2+1 \\
 \epsilon _o^2,\alpha _o,z \\
\end{array}
\right]\, d\gamma.
\end{align}
By taking the Maclaurin series expansion of the complementary error function 
\cite[Eq. (14.47)]{yang2016engineering} 
and utilizing 
\cite[Eq. (2.2.24.2)]{prudnikov1986integrals}, \(\mathcal{C}_b\) is derived as
\begin{align}
\label{ra2}
  \mathcal{C}_b &= \frac{\sqrt{\pi } s^{\alpha _o+z-1}}{ (2 \pi )^{1-s}}
    G_{ s+2, 3 s+1}^{ 3 s ,2 } \left [\frac{\Lambda_0 \text{   }   s^{-2 s}\gamma _{\text{th}} }{\mu _{o_s}\text{}\mathcal{D} _{\text{i}}}   \Bigg| 
\begin{array}{cc}
 1 , \mathcal{T}_1 \\
 \mathcal{T}_2 , 0 \\
\end{array}
 \right]  \nonumber\\
 &-2\sum _{j=1}^{\infty } \text{} 
 \frac{(-1)^j}{(2 j+1) j!}\left(\mathcal{D} _{\text{i}}\right)^{j+\frac{1}{2}}
 \frac{(\gamma _{\text{th}})^{j+\frac{1}{2}} s^{\alpha _o+z-1}}{(2 \pi )^{1-s}} \nonumber\\
 & \times G_{ s+2, 3 s+1}^{ 3 s ,2 } \left [\frac{\Lambda_0\gamma _{\text{th}} s^{-2 s}}{\mu _{o_s}\text{}\mathcal{D} _{\text{i}}}   \Bigg| 
\begin{array}{cc}
\frac{1}{2}-j , \mathcal{T}_1 \\
 \mathcal{T}_2 , -\frac{1}{2}-j \\
\end{array}
 \right].
\end{align}
\setcounter{eqnback}{\value{equation}}
\setcounter{equation}{39}
\begin{figure*}
\begin{align}
\label{count}
    P_{e}^H(\gamma_{th},\gamma_{th}^r)=&\frac{\sum _{\text{i}=1}^{\mathit{n}} \sum _{z=1}^{\beta _o}\Bigg\{\mathcal{M}_1\big[\mathcal{C}_c+ \mathcal{C}_d\big] G_{ s+1, 3 s+1}^{ 3 s ,1 } \left [ \frac{\mathcal{Z} \gamma_{th}}{\mu _{o_s}}  \Bigg| 
\begin{array}{cc}
 1 ,& \mathcal{T}_1 \\
 \mathcal{T}_2 ,& 0 \\
\end{array}
 \right]+\mathcal{M}_2\big[\mathcal{C}_a+ \mathcal{C}_b\big]\Bigg\}}{1- \mathcal{M}_3 \sum _{z=1}^{\beta _o} (\gamma _{\text{th}}) ^{\mathit{g}_t}w_{z}G_{ s+1, 3 s+1}^{ 3 s ,1 } \left [\frac{\mathcal{Z}\gamma _{\text{th}} }{\mu _{o_s}}  \Bigg| 
\begin{array}{cc}
 1 ,& \mathcal{T}_1 \\
 \mathcal{T}_2 ,& 0 \\
\end{array}
 \right] G_{2,3}^{2,1}\left[\chi_3 \left(\frac{\gamma _{\text{th}} }{\overline{\gamma }_t N_t}\right)^{\frac{\alpha_t }{2}}\Bigg| 
\begin{array}{c}
 1-\frac{2 \mathit{g}_t}{\alpha_t },1 \\
 0,\chi_2,-\frac{2 \mathit{g}_t}{\alpha_t } \\
\end{array}
\right]}
\nonumber
\\
\times &\Bigg\{1-2\sum _{\text{i}=1}^{\mathit{n}} \frac{\mathcal{A}}{2 \Gamma  \left(v_r+1\right) \psi _r^{-\frac{1}{2} \left(v_r-1\right)}} \big[\mathcal{C}_e+ \mathcal{C}_f\big]\Bigg\}+\sum _{\text{i}=1}^{\mathit{n}} \frac{\mathcal{A}}{2 \Gamma  \left(v_r+1\right) \psi _r^{-\frac{1}{2} \left(v_r-1\right)}} \big[\mathcal{C}_e+ \mathcal{C}_f\big].
\end{align}
\hrulefill
\end{figure*}
Now, substituting \eqref{pdf t} into \eqref{ber}, \(P_t(\gamma _{\text{th}})\) can be derived as
\setcounter{eqnback}{\value{equation}}
\setcounter{equation}{29}
\begin{align}
\label{ber t}
     P_t (\gamma _{\text{th}})&=\text{}\sum _{\text{i}=1}^{\mathit{n}} \frac{\mathcal{A} \chi_1}{\alpha_t  \left(N_t \overline{\gamma }_t\right) ^{\mathit{g}_t}}\text{}\int _{\gamma _{\text{th}}}^{\infty }\text{erfc}\left(\sqrt{\mathcal{D} _{\text{i}}\gamma }\right)\text{}\gamma ^{\mathit{g}_t-1}\text{}\nonumber\\
    & \times \Gamma \left[\chi_2,\chi_3 \bigg(\frac{\gamma }{ N_t \overline{\gamma }_t}\bigg)^{\alpha_t /2}\right]d\gamma
    \nonumber
    \\
    &= \sum _{\text{i}=1}^{\mathit{n}} \frac{\mathcal{A} \chi_1}{2  \left(N_t \overline{\gamma }_t\right) ^{\mathit{g}_t}} \left(\mathcal{C}_c+ \mathcal{C}_d\right).
\end{align}
The integral term \( \mathcal{C}_c\) is expressed as
\begin{align}
\mathcal{C}_c&=\int _0^{\infty }\text{erfc}\left(\sqrt{\mathcal{D} _{\text{i}}\gamma }\right) \gamma ^{\mathit{g}_t-1} \Gamma \left[\chi_2,\frac{\chi_3 \gamma^{\frac{\alpha_t}{2}}}{(N_t \overline{\gamma }_t)^{\frac{\alpha_t}{2}}}\right]d\gamma.
\end{align}
Converting the upper incomplete Gamma function into a Meijer's $G$ function by \cite[Eq.(8.4.16.2)]{prudnikov1986integrals} and applying the same formula as used in \eqref{ra1},  \( \mathcal{C}_c\) is derived as
\begin{align}
\label{ra3}
   \mathcal{C}_c = \frac{2^{\chi_2}\alpha_t ^{\mathit{g}_t}\mathcal{D} _{\text{i}}^{-\mathit{g}_t}}{(\sqrt{2\pi })^{\alpha_t +1}}G_{2 \alpha_t +2,\alpha_t +4}^{4,2 \alpha_t }\left[\frac{\alpha_t ^{\alpha_t } \chi_3^2}{4N_t \overline{\gamma }_t\mathcal{D} _{\text{i}} ^{\alpha_t }}\Bigg| 
\begin{array}{c}
 \mathcal{T}_3 \\
 \mathcal{T}_4 \\
\end{array}
\right],
\end{align}
where \( \mathcal{T}_3=\left\{\frac{1-\mathit{g}_t}{\alpha_t }\text{},\text{...},1-\frac{\mathit{g}_t}{\alpha_t },\frac{\frac{1}{2}-\mathit{g}_t}{\alpha_t },\text{...}.,\frac{\alpha_t -\mathit{g}_t-\frac{1}{2}}{\alpha_t },\frac{1}{2},1\right\}\) and \(\mathcal{T}_4=\left\{0\text{},\frac{1}{2},\frac{\chi_2}{2},\frac{1}{2} \left(\chi_2+1\right),-\frac{\mathit{g}_t}{\alpha_t },\text{...}.,1-\frac{\mathit{g}_t+1}{\alpha_t }\right\}\).
\\
The integral term $\mathcal{C}_d$ is obtained as
\begin{align}
     \mathcal{C}_d&=\int _0^{\gamma _{\text{th}}}\text{erfc}\left(\sqrt{\mathcal{D} _{\text{i}}\gamma }\right)\gamma ^{\mathit{g}_t-1}  \Gamma \left[\chi_2,\frac{\chi_3 \gamma^{\frac{\alpha_t}{2}}}{(N_t \overline{\gamma }_t)^{\frac{\alpha_t}{2}}}\right]d\gamma.
\end{align}
By converting the upper incomplete Gamma function into a Meijer's $G$ function as in \eqref{ra3} and then utilizing the same identity as utilized in \eqref{ra2}, \(\mathcal{C}_d\) is derived as
\begin{align}
\label{ra4}
\mathcal{C}_d&= \frac{2^{\chi_2-\frac{1}{2}}}{  \sqrt{2 \pi }\alpha_t}\Bigg\{\frac{1}{\gamma _{\text{th}}^{-\mathit{g}_t}}G_{\alpha_t +2,\alpha_t +4}^{4,\alpha_t }\left[\frac{\chi_3^2 (\gamma _{\text{th}})^{\alpha_t }}{4 \left(N_t \overline{\gamma }_t\right) ^{\alpha_t }}\Bigg| 
\begin{array}{c}
 \mathcal{T}_5 \\
 \mathcal{T}_6 \\
\end{array}
\right]\nonumber \\
&-\frac{2}{\sqrt{\pi }}\sum _{j=1}^{\infty } \frac{(-1)^{j}\mathcal{D} _{\text{i}}^{j+\frac{1}{2}}\gamma _{\text{th}}^{\left(j+\frac{1}{2}+\mathit{g}_t\right)}}{(2 j+1) j!} 
\nonumber
\\
&
\times G_{\alpha_t +2,\alpha_t +4}^{4,\alpha_t }\left[\frac{\chi_3^2 (\gamma _{\text{th}})^{\alpha_t }}{4 \left(N_t \overline{\gamma }_t\right)^{\alpha_t }}\Bigg| 
\begin{array}{c}
 \mathcal{T}_7 \\
 \mathcal{T}_8 \\
\end{array}
\right]\Bigg\},
\end{align}
where $\mathcal{T}_5=\left\{\frac{1-\mathit{g}_t}{\alpha_t },\text{...},1-\frac{\mathit{g}_t}{\alpha_t },\frac{1}{2},1\right\}$, $\mathcal{T}_6=\left\{0,\frac{1}{2},\frac{\chi_2}{2},\frac{1}{2} \left(\chi_2+1\right),-\frac{\mathit{g}_t}{\alpha_t },\text{...}.,1-\frac{\mathit{g}_t+1}{\alpha_t }\right\}$, $\mathcal{T}_7=\left\{\frac{\frac{1}{2} (2 j+1)-\mathit{g}_t+1}{\alpha_t },\text{...},1-\frac{\frac{1}{2} (2 j+1)+\mathit{g}_t}{\alpha_t },\frac{1}{2},1\right\}$, and $\mathcal{T}_8=\left\{0,\frac{1}{2},\frac{\chi_2}{2},\frac{1}{2} \left(\chi_2+1\right),-\frac{\frac{1}{2} (2 j+1)+\mathit{g}_t}{\alpha_t },\text{...}.,1-\frac{\frac{1}{2} (2 j+1)+\mathit{g}_t+1}{\alpha_t }\right\}$.

\subsection*{Derivation of  \(P_a\):}
\noindent
\(P_a\) can be obtained by substituting \eqref{pdf r} into \eqref{ber} as
\begin{align}
\label{ber r}
P_{a} (\gamma _{\text{th}}^r)&=\sum _{\text{i}=1}^{\mathit{n}} \frac{\mathcal{A}\psi _r^{\frac{v_r-1}{2}}}{2 \Gamma  \left(v_r+1\right) }\int_{\gamma _{\text{th}}^r}^{\infty } \gamma ^{\frac{v_r-1}{2}}e^{-\sqrt{\gamma  \psi _r}} \text{erfc}\left(\sqrt{\mathcal{D} _{\text{i}}\gamma }\right) \, d\gamma
\nonumber
\\
&=\sum _{\text{i}=1}^{\mathit{n}} \frac{\mathcal{A}}{2 \Gamma  \left(v_r+1\right) \psi _r^{-\frac{1}{2} \left(v_r-1\right)}} \left(\mathcal{C}_e+ \mathcal{C}_f\right).
\end{align}

The integral term \( \mathcal{C}_e\) is expressed as
\begin{align}
     \mathcal{C}_e=\int_0^{\infty } \gamma ^{\frac{1}{2} \left(v_r-1\right)} e^{-\sqrt{\gamma  \psi _r}} \text{erfc}\left(\sqrt{\mathcal{D} _{\text{i}}\gamma }\right) \, d\gamma.
\end{align}
By converting exponential function into a Meijer's $G$ function \cite[Eq.(8.4.3.1)]{prudnikov1986integrals} and using the same formula as used in \eqref{ra1} and \eqref{ra3}, \( \mathcal{C}_e\) is obtained as
\begin{align}
\label{yo}
\mathcal{C}_e=\frac{\mathcal{D} _{\text{i}}^{-\frac{1}{2} \left(v_r+1\right)}}{\sqrt{\pi }}G_{2,3}^{2,2}\left[\frac{\psi _r}{4\mathcal{D} _{\text{i}}}\Bigg| 
\begin{array}{c}
\frac{1}{2}(1-v_r) ,-\frac{v_r}{2}\\
0, \frac{1}{2}, 1\\
\end{array}\right].
\end{align}
The integral term
\(\mathcal{C}_f\) is derived as
\begin{align}
     \mathcal{C}_f=\int_{0}^{\gamma _{\text{th}}^r} \gamma ^{\frac{1}{2} \left(v_r-1\right)} e^{-\sqrt{\gamma  \psi _r}} \text{erfc}\left(\sqrt{\mathcal{D} _{\text{i}}\gamma }\right) \, d\gamma.
\end{align}
By converting exponential function into a Meijer's $G$ function as in \eqref{yo} and utilizing the same identity as used in \eqref{ra1} and \eqref{ra4}, 
\(\mathcal{C}_f\) is obtained as
\begin{align}
\mathcal{C}_f&=\frac{\sqrt{2} (\gamma^r _{\text{th}})^{\frac{1}{2} \left(v_r+1\right)}}{\sqrt{2 \pi }}G_{1,3}^{2,1}\left[\frac{\psi _r \gamma _{\text{th}}^r}{4}\bigg|
\begin{array}{c}
1-\frac{1}{2}\left(v_r+1\right)\\
0,\frac{1}{2},-\frac{1}{2}\left(v_r+1\right)\\
\end{array}
\right] \nonumber\\
&-\frac{2}{\pi }\sum _{j=1}^{\infty } \frac{(-1)^{j}\mathcal{D} _{\text{i}}^{j+\frac{1}{2}}(\gamma^r_{\text{th}})^{j+1+\frac{v_r}{2}}}{(2 j+1) j!}\nonumber\\
&\times G_{1,3}^{2,1}\left[\frac{\psi _r \gamma _{\text{th}}^r}{4}\bigg|
\begin{array}{c}
 - \left( j+\frac{v_r}{2}\right) \\
 0,\frac{1}{2},- \left( j+1+\frac{v_r}{2}\right) \\
\end{array}
\right].
\end{align}
Now, substituting Eqs. \eqref{malaga}, \eqref{first hop}, \eqref{ber f1}, and \eqref{ber t} into \eqref{ber hyb h}, we get the ABER expression for the hybrid link. Then, placing \eqref{ber hyb h} and \eqref{ber r} into \eqref{ber1}, the analytical expression of ABER due to the proposed model is obtained as \eqref{count}, where $\mathcal{M}_1=\frac{\mathcal{A}w_{z} \chi_1\Omega _o \epsilon _o^2}{2  \left(N_t \overline{\gamma }_t\right) ^{\mathit{g}_t}2^{2 s-1} \pi ^{s-1}} $, $\mathcal{M}_2=\frac{\mathcal{A} \Omega _o \epsilon _o^2 v_z}{\sqrt{\pi } 2^{s}}$,
and 
$\mathcal{M}_3=\frac{\chi_1 \Omega _o \epsilon _o^2}{2^{2 s-1} \pi ^{s-1}\alpha_t  \left(N_t \overline{\gamma }_t\right)^{\mathit{g}_t}} $.
\subsubsection{Soft Switching}
For the soft switching case, the analytical expression of ABER due to the proposed model can be obtained mathematically as
\setcounter{eqnback}{\value{equation}}
\setcounter{equation}{40}
\begin{align}
\label{ber22}
   P_\mathbf{e}^S=\text{} P_{\text{hyb}}^S-2P_a P_{\text{hyb}}^S+P_a.
\end{align}
The analytical expression of ABER due to the hybrid FSO/THz link can be written as
\begin{align}
\label{soft ber hybrid}
P_{\text{hyb}}^{S}(\gamma_{th}^u,\gamma_{th}^l,\gamma_{th}^t) =&\frac{1}{1 - F_{\gamma_s}}  \Bigl\{ P_o (\gamma_{\text{th}}^u) + F_{\gamma_o}(\gamma_{\text{th}}^u,\gamma_{\text{th}}^l) P_t(\gamma_{\text{th}}^t)
\nonumber
\\
+& \frac{[P_o (\gamma_{\text{th}}^l) - P_o (\gamma_{\text{th}}^u) ][1 - F_{\gamma_o}(\gamma_{\text{th}}^u)]}{F_{\gamma_o}(\gamma_{\text{th}}^l) - F_{\gamma_o}(\gamma_{\text{th}}^u) + 1} \Bigl\},
\end{align}
where \(F_{\gamma _o} (\gamma _{\text{th}}^u)\), \(F_{\gamma _o} (\gamma _{\text{th}}^l)\), \(F_{\gamma _s}\), \(P_o (\gamma _{\text{th}}^u)\), \(P_o (\gamma _{\text{th}}^l)\), \(P_t (\gamma _{\text{th}}^t)\), and \(F_{\gamma _o}\left(\gamma _{\text{th}}^u, \gamma _{\text{th}}^l\right)\) are obtained from  \eqref{malaga}, \eqref{soft hybrid}, \eqref{ber f1}, \eqref{ber t}, and  \eqref{soft m}, respectively, by substituting the appropriate thresholds. Finally, by substituting \eqref{soft ber hybrid} and \eqref{ber r} into \eqref{ber22}, the CFE of ABER is obtained as shown in \eqref{big_equation soft ber}. Where, $\mathcal{M}_4=  \frac{\Omega_o\,\epsilon_o^2}{2^{2s-1}\,\pi^{s-1}}$, $\mathcal{M}_5=\frac{\chi_1\mathcal{A}\Omega_ov_z\epsilon_o^2(\gamma_{th}^t)^{g_t} }{\sqrt\pi\alpha_t\,(N_t\overline\gamma_t)^{g_t}\,2^s}$, and $\mathcal{M}_6=  \frac{\mathcal{A}\,\chi_1}{2\,(N_t\overline\gamma_t)^{g_t}}$.
\setcounter{eqnback}{\value{equation}}
\setcounter{equation}{42}
\begin{figure*}
\begin{align}
\label{big_equation soft ber}
&P_{\text{e}}^S(\gamma_{th}^u,\gamma_{th}^l,\gamma_{th}^t,\gamma_{th}^r)
=
\Biggl\{
1
-
 \mathcal{M}_4
  \sum_{z=1}^{\beta_o} w_z\,
  G_{s+1,3s+1}^{3s,1}\left[\tfrac{\mathcal{Z}\gamma_{th}^l}{\mu_{o_s}}
    \Bigm|\begin{matrix}1,&\mathcal{T}_1\\\mathcal{T}_2,&0\end{matrix}\right]
  + G_{s+1,3s+1}^{3s,1}\left[\tfrac{\mathcal{Z}\gamma_{th}^u}{\mu_{o_s}}
\Bigm|\begin{matrix}1,&\mathcal{T}_1\\\mathcal{T}_2,&0\end{matrix}\right]
\Biggl\}^{-1}
\sum_{z=1}^{\beta_o}\sum_{i=1}^n\mathcal{M}_5  
\nonumber\\
\quad
& \times \bigl[\mathcal{C}_a
         +\mathcal{C}_b\bigr]
G_{2,3}^{2,1}\left[\,
  \chi_3\bigl(\tfrac{\gamma_{th}^t}{N_t\overline\gamma_t}\bigr)^{\frac{\alpha_t}{2}}
  \Bigm|\begin{matrix}
    1-\tfrac{2g_t}{\alpha_t},\,1\\
    0,\,\chi_2,\,-\tfrac{2g_t}{\alpha_t}
  \end{matrix}\right]
+F_{\gamma_o}(\gamma_{\text{th}}^u,\gamma_{\text{th}}^l)\sum_{i=1}^n\mathcal{M}_6\bigl[\mathcal{C}_c+\mathcal{C}_d\bigr] \Bigg\{1- 2\sum _{\text{i}=1}^{\mathit{n}} \frac{\mathcal{A}\big[\mathcal{C}_e+ \mathcal{C}_f\big]}{2 \Gamma  \left(v_r+1\right) \psi _r^{-\frac{1}{2} \left(v_r-1\right)}} 
\nonumber\\
\qquad\quad
& \times 
  \frac{
    1
    -
   \mathcal{M}_4
    \sum_{z=1}^{\beta_o} w_z\,
    G_{s+1,3s+1}^{3s,1}\left[\tfrac{\mathcal{Z}\,\gamma_{\text{th}}^u}{\mu_{o_s}}
    \Bigm|\begin{matrix}1,&\mathcal{T}_1\\\mathcal{T}_2,&0\end{matrix}\right]
  }
  {\mathcal{M}_4
    \sum_{z=1}^{\beta_o} w_z\,
    G_{s+1,3s+1}^{3s,1}\left[\tfrac{\mathcal{Z}\,\gamma_{\text{th}}^l}{\mu_{o_s}}
      \Bigm|\begin{matrix}1,&\mathcal{T}_1\\\mathcal{T}_2,&0\end{matrix}\right]
    -
    G_{s+1,3s+1}^{3s,1}\left[\tfrac{\mathcal{Z}\,\gamma_{\text{th}}^u}{\mu_{o_s}}
      \Bigm|\begin{matrix}1,&\mathcal{T}_1\\\mathcal{T}_2,&0\end{matrix}\right]
    +1
  }+\sum _{\text{i}=1}^{\mathit{n}} \frac{\mathcal{A}\psi _r^{\frac{1}{2} \left(v_r-1\right)}}{2 \Gamma  \left(v_r+1\right) } \big[\mathcal{C}_e+ \mathcal{C}_f\big]\Bigg\}.
\end{align}
\hrulefill
\end{figure*}
\subsection{Significance of the Derived Expressions}
The main objective of this work was to investigate the system performance of the proposed network by utilizing the physical properties of the FSO/THz–RF propagation medium. To achieve this goal, all system parameters were analytically related to two key performance metrics such as OP and ABER. To the best of our knowledge, this study was the first to consider an aerial RIS in such a network to improve overall system performance. Therefore, the derived analytical expressions can be regarded as entirely novel. 

Since the considered channels are generalized, various existing models can be derived as special cases of the proposed framework. For instance, by setting $\rho_0=1$, $\mathcal{J}_p=0$, and $\text{ó}=1$ in Eqs. (\ref{eq 36}), (\ref{eq 37}), (\ref{count}), and (\ref{big_equation soft ber}), the model simplifies to the GG/$\alpha-\mu$-Rician mixed model. Furthermore, by setting $J_k=0$ in our derived mathematical expressions, the proposed model reduces to the Málaga/$\alpha-\mu$-Rayleigh dual-hop hybrid model. Therefore, the derived analytical expressions are not only novel but also versatile, enabling the analysis of a wide range of other mixed fading scenarios.
\subsection{Asymptotic Analysis}
Asymptotic analysis is a mathematical approach to approximating system performance at high SNR levels by simplifying complex expressions. It provides valuable insights into the system's behavior under ideal conditions.

\subsubsection{Asymptotic Outage Probability} 
For the hard switching approach, by applying the identity \cite[Eq.~(8.2.2.14)]{prudnikov1986integrals} and subsequently using \cite[Eq.~(30)]{ansari2015performance}, the Meijer’s G term is expanded. The asymptotic outage probability of the hybrid link $(\Bar{\gamma}_t,\mu_{o_s}\to \infty)$ can be obtained as
\begin{align}
\label{ahy}
    P_{out}^{hyb}(\infty)&=\sum _{z=1}^{\beta _o}\sum _{i=1}^{3s}\sum _{l=1}^{2}\frac{\chi_1w_{z}\Omega _o \epsilon _o^2 (\gamma_{th}) ^{\mathit{g}_t}}{\alpha_t  \left(N_t \overline{\gamma }_t\right)^{\mathit{g}_t}2^{2 s-1} \pi ^{s-1}}\Bigl(\frac{\mu _{o_s}}{\mathcal{Z} \gamma_{th}}\Bigl)^{M_{i}-1}
 \nonumber 
 \\
 &\times\Biggl\{
 \frac{(\overline{\gamma }_t N_t)^{{\frac{\alpha_t}{2}}}}{ \chi_3(\gamma_{th})^{\frac{\alpha_t}{2}}}
 \Biggl\}^{B_{l}-1} \frac{\prod_{j=1,j\neq i}^{3s} \Gamma(M_{i}-M_{j})}{\Gamma(1+M_{3s+1}-M_{i})}
 \nonumber
 \\
 &\times 
 \frac{\Gamma(1+N_{1}-M_i)\prod_{k=1,k\neq l}^{2} \Gamma(B_{l}-B_{k})}{\prod_{j=2}^{s+1} \Gamma(M_{i}-N_{j})\Gamma(1+B_{3}-B_{l})}
 \nonumber
 \\
 &\times 
 \frac{\Gamma(1+D_{1}-B_l)}{\Gamma(B_{l}-D_{2})},
\end{align}
where $M=\begin{bmatrix}
    1-\mathcal{T}_{2} & 1
\end{bmatrix}$, $N=\begin{bmatrix}
    0 & 1-\mathcal{T}_{1}
\end{bmatrix}$, $B=\begin{bmatrix}
    1 & 1-\chi_{2} & 1+\frac{2g_t}{\alpha_t}
\end{bmatrix}$, and $D=\begin{bmatrix}
    \frac{2g_t}{\alpha_t} & 0
\end{bmatrix}$. Applying the same formulas as used in \eqref{ahy}, the asymptotic outage probability of the RF link $(\Bar{\gamma}_r \to \infty)$ can be written as
\begin{align}
     \label{arf}P_{out}^{rf}(\infty)&=\sum_{g=1}^{2}\frac{(\psi _r \gamma_{th})^{-\frac{(R_{g}-1)}{2}}}{\Gamma\left(v_r+1\right)}\frac{\prod_{h=1;h \neq g}^{2} \Gamma(R_{g}-R_{h})}{\Gamma(R_{g})},
\end{align}
where $R=\begin{bmatrix}
    -v_r & 1
\end{bmatrix}$. Therefore, the asymptotic OP for the dual-hop mixed network can be expressed as 
\begin{align}
    P_{out}^{H}(\infty)= P_{out}^{hyb}(\infty) +  P_{out}^{rf}(\infty). 
\end{align}
Now, we consider two distinct cases. When the hybrid link dominates the proposed network, the first hop becomes the determining factor in system performance. This condition arises when the hybrid link experiences more severe impairments than the RF link, primarily due to high pointing errors, strong atmospheric turbulence, and deep fading. Accordingly, the asymptotic expression of the OP can be written as $P_{out}^{H}(\infty)=(G_c^{h} \Bar{\gamma})^{-G_d^{h}}$, where $G_c^{h}$ and $G_d^{h}$ denote the coding gain and diversity order, respectively. The corresponding diversity order is then derived as
\begin{align}
    G_d^h=min\biggl\{\frac{\epsilon _o^2}{s},\frac{\alpha _o}{s},\frac{z}{s},\frac{\alpha_t\mu_tN_t}{2}\biggl\}.
\end{align}
In the second case, when the RF link dominates, the second hop becomes the primary performance-limiting factor. This scenario occurs when the RF link experiences more severe degradation than the hybrid link, mainly due to strong fading effects in the RF channel. Under this condition, the asymptotic expression of the OP can be expressed as $P_{out}^{H,\infty}=(G_c^{r} \Bar{\gamma})^{-G_d^{r}}$, where $G_c^{r}$ and $G_d^{r}$ represent the coding gain and diversity order of the RF link, respectively. Consequently, the diversity order is obtained as 
\begin{align}
    G_d^r=\frac{v_r+1}{2}.  
\end{align}
Similarly, applying the same identities as utilized in \eqref{ahy} and \eqref{arf}, all the Meijer's $G$ terms in \eqref{eq 37} are expanded. By substituting the obtained terms in \eqref{eq 37}, the asymptotic OP for soft switching is derived. 

\subsubsection{Asymptotic Average Bit Error Rate}
By utilizing similar procedure as \eqref{ahy} and \eqref{arf}, all the Meijer’s $G$ terms are expanded. Now, substituting the expanded terms appropriately in \eqref{count} and \eqref{big_equation soft ber}, the asymptotic ABER for hard and soft switching schemes are obtained.
\section{Numerical Results}
 \label{sec4}
This section presents a graphical representation of the numerical results obtained from the derived analytical expressions of OP and ABER. The key objective of this investigation is to analyze the impact of various parameters ($N_{t}$, $N_{r}$, $\alpha_o$, $\beta_o$, $\alpha_t$, $\beta_t$, $\epsilon_0$, $\epsilon_t$, $\overline{\gamma}_{t}$, $\overline{\gamma}_{r}$, $\mathcal{R}_1$, $\mathcal{R}_2$, $\mathcal{H}$) on the system performance for the proposed RIS-assisted dual hop network. For a more comprehensive analysis, we set the parameters according to Table \ref{simt}.
\begin{table*}[!ht]
\centering
\caption{Simulation Parameter Settings \cite{ibrahim,rakib,ris}}
\begin{tabular}{l l l l l l}
\hline
\multicolumn{2}{c}{\textbf{FSO Link}} & \multicolumn{2}{c}{\textbf{THz Link}} & \multicolumn{2}{c}{\textbf{RF Link}} \\ \hline
\textit{\textbf{Parameters}} & \textit{\textbf{Value}} & \textit{\textbf{Parameters}} & \textit{\textbf{Value}} & \textit{\textbf{Parameters}} & \textit{\textbf{Value}} \\
Strong Turbulence ($\alpha_o,\beta_o$)      & $2.296, 2$   & Fading Parameters ($\alpha_t,\mu_t$)   & ($2,2$), ($2,3$)   & Horizontal Distance (km) ($\mathcal{R}_1,\mathcal{R}_2$)   & ($3,3$), ($3,3.5$), ($1,1$) \\ 
Moderate Turbulence ($\alpha_o,\beta_o$)     & $4.2, 3$     & Pointing Error  ($\epsilon_t$)     & $1, 1.5$         & UAV’s altitude (km)   ($\mathcal{H}$)     & $1, 2, 3$               \\ 
Weak Turbulence  ($\alpha_o,\beta_o$)       & $8, 4$       & Number of Antennas ($N_t$)   & $2, 3$           & Number of Elements in RIS ($N_r$) & $1, 3, 5$               \\
   Average SNR ($\overline{\gamma }_o$)     & $35$ dB  & Average SNR ($\overline{\gamma }_t$)     & $25$ dB    & Average SNR ($\overline{\gamma }_r$)     & $40$ dB  \\            
 Threshold SNRs ($\gamma_{th}^l,\gamma_{th}^u$)     & $15,25$ dB   &  Threshold SNR ($\gamma_{th}^t$)     & $20$ dB    & Threshold SNR ($\gamma_{th}^r$)     & $60$ dB      \\
 Pointing Error  ($\epsilon_o$)        & $1, 6.7$  &      &    &     &       \\

     \hline
     \label{simt}
\end{tabular}
\end{table*}
These graphical representations are further validated through MC simulations, which utilize random variable generation in MATLAB to provide a statistically significant assessment of system performance by averaging $10^6$ random samples for each channel. Finally, asymptotic analysis are performed to show the result in the high region.
\subsection{Impact of FSO parameters}
\begin{figure}[!ht]
\vspace{0mm}
\centerline{\includegraphics[width=0.35\textwidth,angle=0]{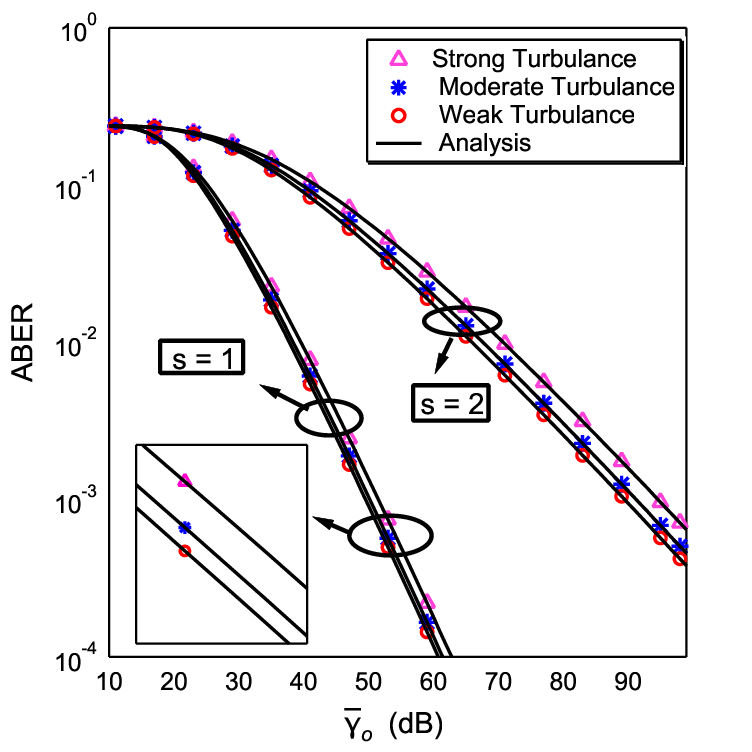}}
    \vspace{0mm}
    \caption{ABER versus \(\overline{\gamma }_o\) for different turbulence conditions and receiver detection techniques.}
    \label{fig 2}
\end{figure}

Fig. \ref{fig 2} illustrates the impact of atmospheric turbulence parameters ($\alpha_o$, $\beta_o$) on system performance.
It is observed that ABER performance improves significantly under weak turbulence compared to moderate and strong turbulence conditions. This improvement is attributed to the fact that higher turbulence parameters correspond to reduced turbulence severity, leading to diminished scattering, lower signal fading, and fewer optical distortions. As a result, a greater portion of the transmitted optical signal reaches the receiver with improved integrity. Furthermore, the figure provides a comparative assessment of two widely used detection techniques, namely HD and IM/DD. The results clearly indicate that HD detection substantially outperforms the IM/DD method, as observed by a significant reduction in ABER when the HD scheme is applied at the receiver.
\begin{figure}[!ht]
\vspace{0mm}
\centerline{\includegraphics[width=0.35\textwidth,angle=0]{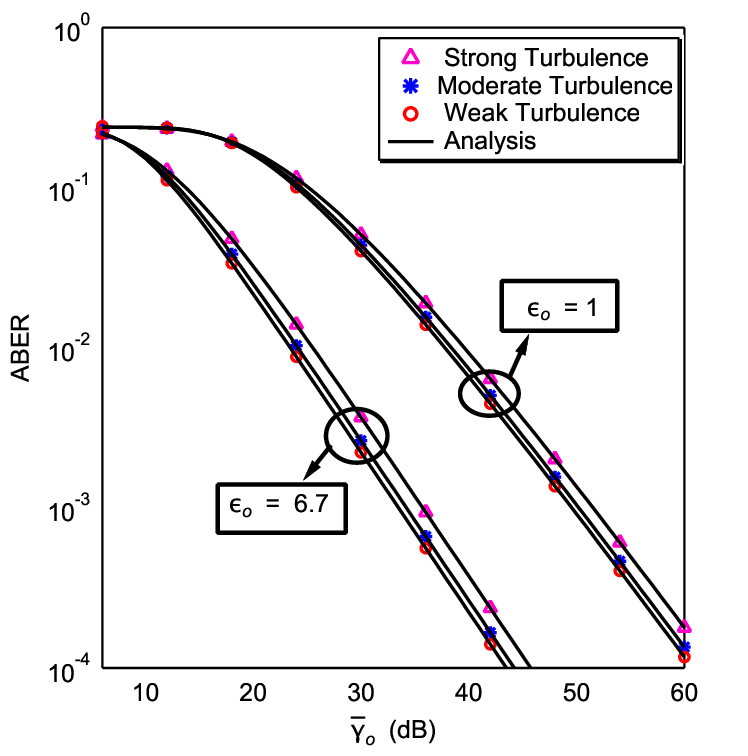}}
    \caption{ABER versus \(\overline{\gamma }_o\) under different turbulence and pointing errors.}
    \label{fig 3}
\end{figure}
This performance gain is expected, given that HD uses a local oscillator to detect weak signals, enabling extraction of both amplitude and phase information. By analyzing signal pulses over time, HD offers superior adaptability to channel fluctuations, thereby effectively mitigating the detrimental effects of turbulence-induced impairments.

The impact of the pointing error parameter ($\epsilon_o$) on the performance of the optical network is examined in Fig. \ref{fig 3}.
The results indicate a significant reduction in ABER with increasing values of $\epsilon_o$, reflecting a marked improvement in system performance. This enhancement is primarily due to the reduction in misalignment between the transmitted optical beam and the receiver, which effectively mitigates the severity of pointing errors.
\subsection{Impact of THz parameters}
\begin{figure}[!ht]
\vspace{0mm}
\centerline{\includegraphics[width=0.35\textwidth,angle=0]{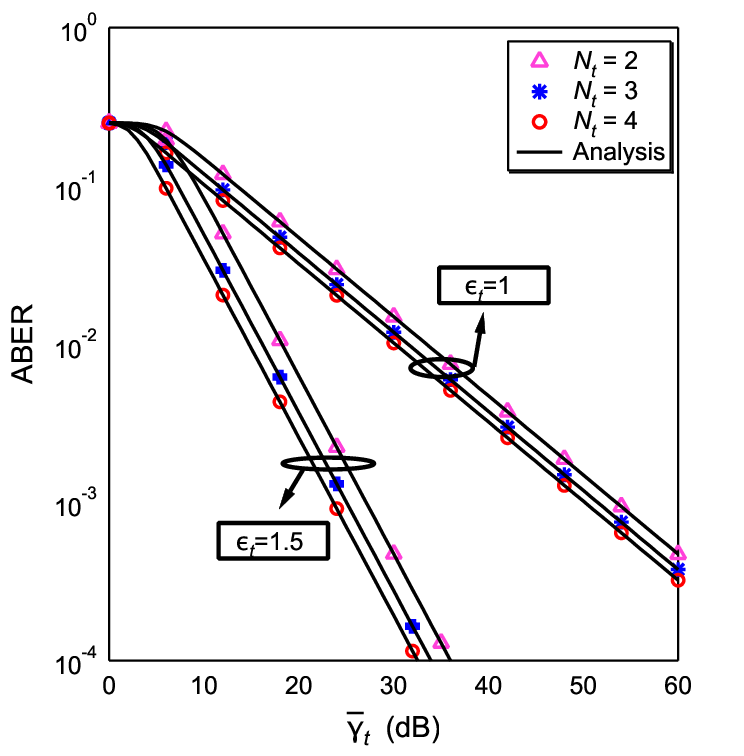}}
    \vspace{0mm}
    \caption{ABER vs \(\overline{\gamma }_t\) under different numbers of antenna and pointing error.}
    \label{fig 4}
\end{figure}

The effect of the number of antennas, $N_t$ in the hybrid network is analyzed by plotting ABER against $\overline{\gamma }_t$ in Fig. \ref{fig 4}.
The figure demonstrates that ABER performance improves as the number of antennas increases. A higher number of antennas at the THz access point allows more of the transmitted signal to be captured. This leads to stronger received signals. As a result, the signal quality improves and the rate of data corruption is reduced. For $N_t = 2$, the ABER is $ 5.93 \times 10^{-2}$ at $20$ dB. Increasing $N_t$ to $4$ at the same SNR reduces the value of ABER to $4.05 \times 10^{-2}$, resulting in a $32.20\%$ overall improvement in system performance. The figure also highlights the effect of the pointing error parameter, $\epsilon_t$ on the THz link. The results show that increasing $\epsilon_t$ from $1$ to $1.5$ significantly reduces the ABER from $9.440 \times 10^{-2}$ to $2.930 \times 10^{-2}$ at 12 dB. This leads to a clear improvement in system performance. A higher value of $\epsilon_t$ reduces the misalignment between the source and the access point. This improves the quality of the received signal.

\begin{figure}[!ht]
\vspace{0mm}
\centerline{\includegraphics[width=0.35\textwidth,angle=0]{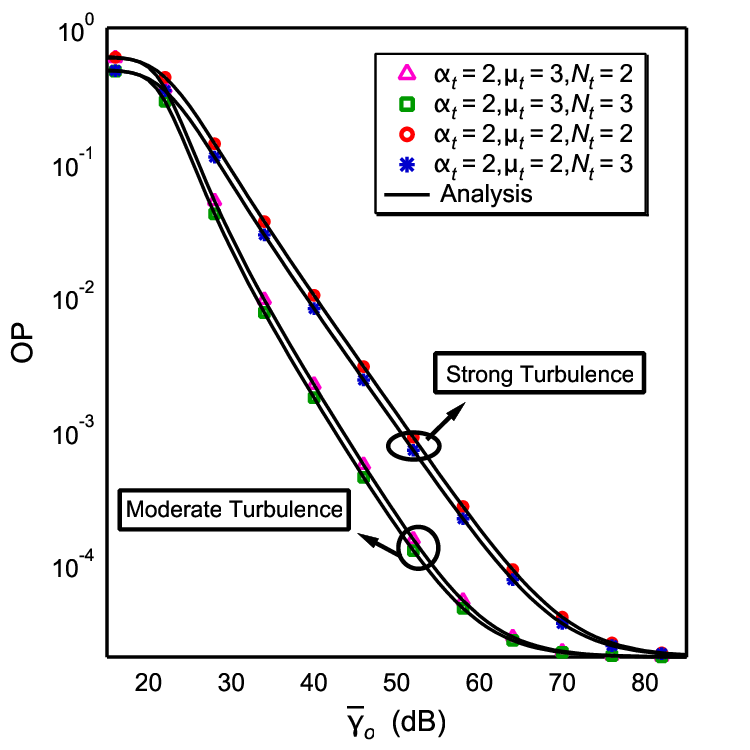}}
    \vspace{0mm}
    \caption{OP versus  \(\overline{\gamma }_o\) for various fading parameters.}
    \label{fig 6}
\end{figure}

To examine the impact of fading severity in the THz link, the OP is plotted against the $\overline{\gamma }_o$ in Fig. \ref{fig 6}.
The results show that OP performance improves when the fading parameters ($\alpha_t$, $\mu_t$) increase from ($2,2$) to ($2,3$). This improvement is expected because higher values of these parameters reduce the severity of fading. As a result, the received signal strength increases. Therefore, as ($\alpha_t$, $\mu_t$) increases, the system exhibits a lower outage probability, indicating better overall performance.
\subsection{Impact of UAV-aided RIS}
\begin{figure}[!ht]
    \centering
\includegraphics[width=0.35\textwidth]{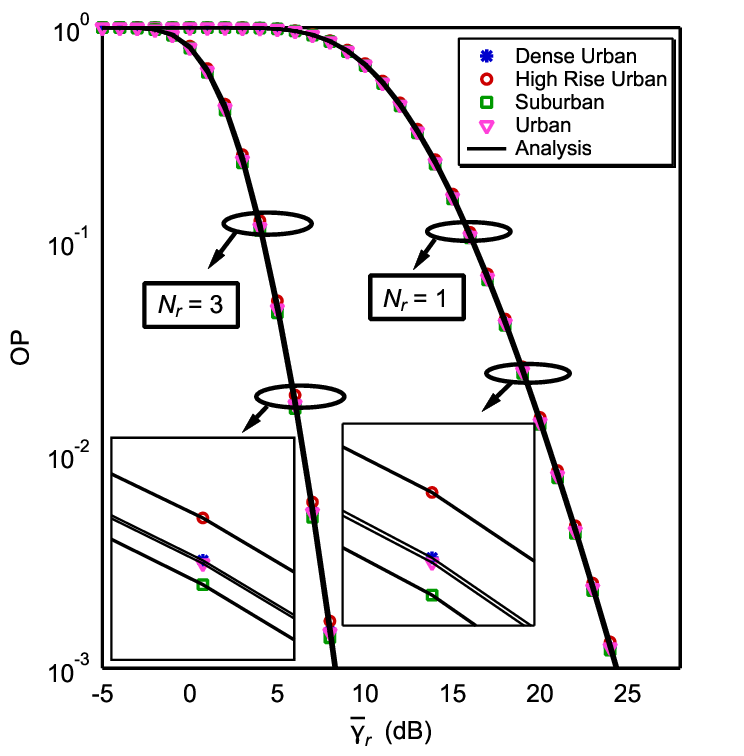}
    \caption{OP versus \(\overline{\gamma }_r\)  in different environments for RIS elements.}
    \label{fig 11}
\end{figure}

The benefit of incorporating RIS in the proposed model is illustrated in Fig. \ref{fig 11}. To evaluate this impact, OP is plotted against \(\overline{\gamma }_r\)  under different communication environments, including suburban, urban, dense urban, and high-rise urban scenarios. The results indicate that OP increases as the environmental density increases, reflecting a significant decline in system performance. 
This trend is expected, as denser environments introduce more signal blockage and scattering. However, the use of RIS can mitigate these adverse effects. By increasing the number of reflecting elements, $N_r$, the RIS can more effectively direct the signal toward the receiver, enhancing signal strength and improving link quality. At $10$ dB, increasing the value of $N_r$ from $1$ to $3$ significantly reduces the OP from $9.8\times10^{-1}$ to $3.8\times10^{-1}$. This represents a $61.22\%$ improvement in performance at relatively low SNR, indicating higher number of reflecting elements enhance signal efficiency by reducing signal power loss.

\begin{figure}[!ht]
    \centering
     \includegraphics[width=0.35\textwidth]{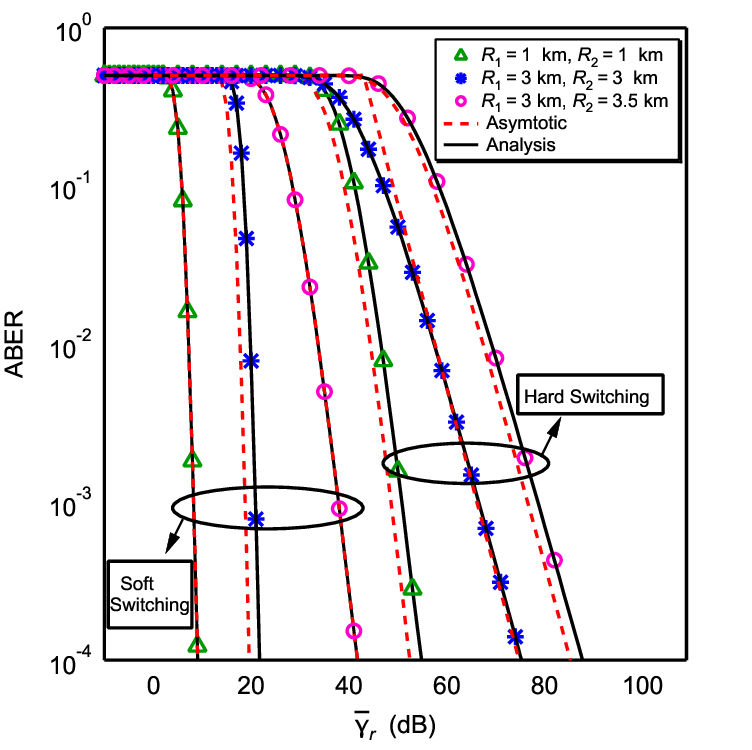}
    \caption{ABER versus \(\overline{\gamma }_r\) for different values of \(\mathcal{R}_1\) and \(\mathcal{R}_2\).}
    \label{fig 8}
\end{figure}

The optimal placement of a UAV-aided RIS is crucial for enhancing the overall performance of wireless communication systems. To evaluate this impact, the ABER is plotted against $\overline{\gamma }_r$ in Fig. \ref{fig 8}. Three distinct scenarios are considered for the analysis: (a) the farthest distance scenario, where the UAV maintains equal but large distances from both the access point and the mobile user (i.e., $\mathcal{R}_1 = \mathcal{R}_2 = 3$ km), (b) the unequal distance scenario, where the UAV is positioned closer to the access point than the mobile user(i.e., $\mathcal{R}_1 = 3$ km, $\mathcal{R}_2 = 3.5$ km), and (c) the closest distance scenario, where the UAV is positioned equidistant at a shorter range from both endpoints (i.e., $\mathcal{R}_1 = \mathcal{R}_2 = 1$ km).
The results indicate that the system exhibits the worst performance when the UAV is placed at unequal distances from the access point and the mobile user. This imbalance causes asymmetric path losses, leading to signal distortion and increased bit error rates. On the other hand, when the UAV maintains equal distances from both nodes, a significant improvement in ABER is observed. This highlights the benefit of symmetric placement, as it ensures balanced signal reflection and minimal path loss, effectively optimizing the RIS-assisted transmission. Moreover, the analysis reveals that the system performs better at shorter distances, even when the distances remain symmetric. In particular, reducing both $\mathcal{R}_1$ and $\mathcal{R}_2$ to $1$ km further lowers the ABER due to decreased signal attenuation and improved SNR at the receiver. Considering hard switching scheme at \(\overline{\gamma }_r\) = $37$ dB, the ABER for  farest and unequal distance is \(4.8 \times 10^{-1}\), while for closest and equal distance scenario, it reduces to \(2.8 \times 10^{-1}\). This comparison demonstrates a $41.67\%$ improvement in ABER when the optimal placement of UAV is implemented.
\begin{figure}[!ht]
    \centering
    \includegraphics[width=0.35\textwidth]{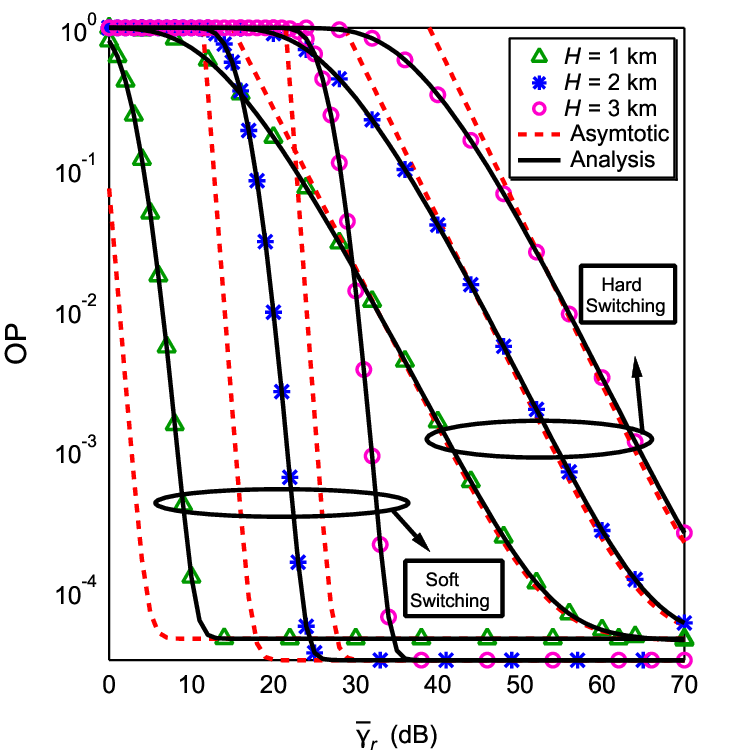}
    \caption{OP versus \(\overline{\gamma }_r\) for different values of \(\mathcal{H}\).}
    \label{fig 9}
\end{figure}

Fig. \ref{fig 9} illustrates the impact of UAV altitude, $\mathcal{H}$ on the OP performance. When  $\mathcal{H} = 3$ km at \(\overline{\gamma }_r\) = $12$ dB, the OP for the hard switching technique is observed to be \(8.6 \times 10^{-1}\). There is a discernible improvement when the altitude is lowered to $1$ km, as the OP drops to \(5.8 \times 10^{-1}\), which is a reduction of $32.56\%$. In comparison, the soft switching technique consistently outperforms hard switching. At an altitude of $3$ km, the OP under soft switching is significantly lower at \(6.7 \times 10^{-1}\), which further declines to \(4.3 \times 10^{-1}\) when the UAV is repositioned to $1$ km, marking a $35.82\%$ improvement. The results demonstrate that as $\mathcal{H}$ increases, the OP also increases, reflecting a noticeable degradation in system performance. This degradation is primarily due to increased path loss, as higher altitudes lead to longer transmission distances and weaker received signal strength, thereby reducing the SNR. At lower altitudes, the UAV is more likely to sustain a LoS connection, which supports stronger and more stable communication links. However, as the altitude rises, the probability of encountering obstacles also increases, potentially causing non-LoS conditions and further deteriorating the link quality.
\subsection{Impact of Modulation and Switching Technique}
\begin{figure}[!ht]
    \centering
\includegraphics[width=0.35\textwidth]{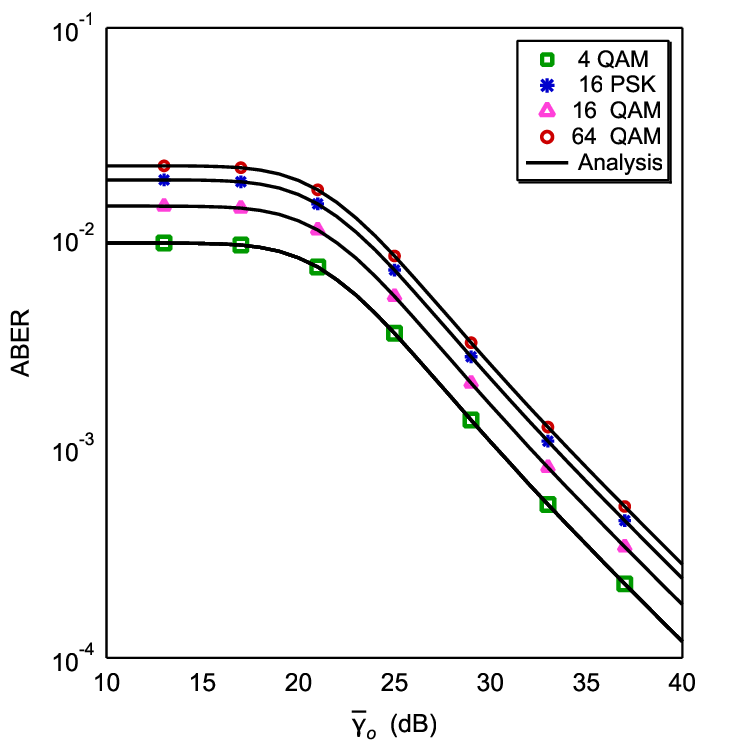}
    \caption{ABER versus \(\overline{\gamma }_o\) for various modulation schemes.}
    \label{fig 12}
\end{figure}

In Fig. \ref{fig 12}, the ABER performance is evaluated under various modulation schemes as shown in Table \ref{mod tab}. As the modulation schemes transition from simpler to more complex, the ABER typically increases. For instance, moving from $4$-Quadrature Amplitude Modulation ($4$-QAM) to a more intricate scheme like $64$-QAM results in a degraded ABER performance. This is because while complex schemes can transmit more data per symbol, they are also more vulnerable to noise and errors. For the same constellation order, $M$-QAM shows improved performance as compared to the M-Phase Shift Keying (PSK) scheme. From the figure, it is observed that $16$-QAM offers about $25\%$ improvement over $16$-PSK at $13$ dB. This improvement arises because $M$-PSK modulates only the phase of the constellation points, whereas $M$-QAM modulates both the amplitude and phase. However, increasing the QAM constellation order from $16$ to $64$ increases the ABER value from $ 1.416 \times 10^{-2}$ to $2.203 \times 10^{-2}$, leading to a $35.72\%$ decrease in system performance.
\begin{figure}[!ht]
    \centering
\includegraphics[width=0.35\textwidth]{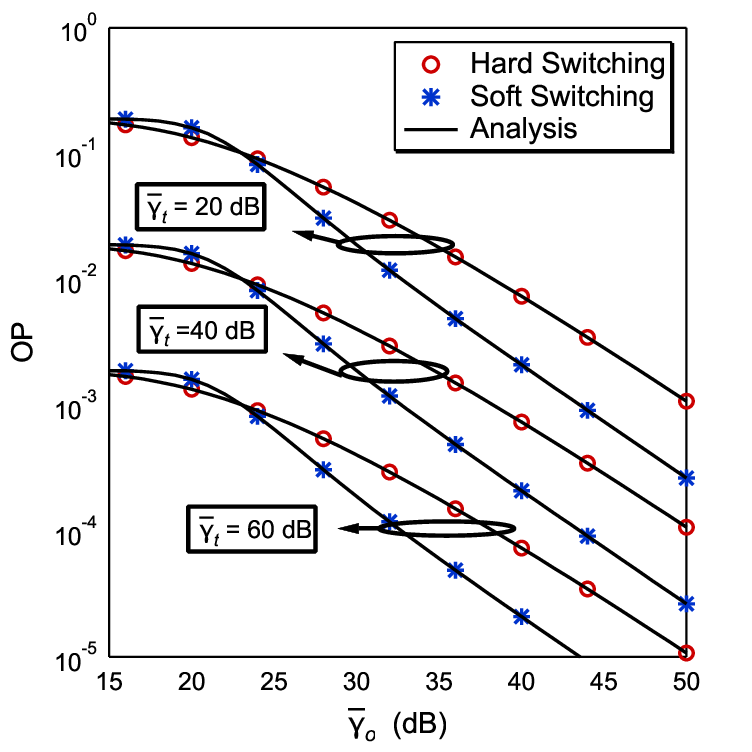}
    \caption{OP versus \(\overline{\gamma }_o\) for hard and soft switching under \(\overline{\gamma }_t\).}
    \label{fig 7}
\end{figure}

To analyze the impact of switching techniques at the receiver of the hybrid network, Fig. \ref{fig 7} illustrates the system performance under different switching condition. The results clearly demonstrate that soft switching provides better performance compared to hard switching. This improvement is expected, as soft switching does not rely on a fixed threshold to select between links. Instead, it dynamically adapts to real-time channel conditions, enabling the system to prioritize the FSO link as long as its performance remains acceptable. This approach avoids premature switching to the THz link and ensures more consistent use of the high-capacity FSO channel. As a result, soft switching reduces service interruptions and improves overall signal quality. Based on these observations, the proposed model strongly recommends that network engineers implement soft switching at the access point to achieve the best system performance. Fig. \ref{fig 7} further examines the impact of the average SNR of the THz link, $\overline{\gamma}_t$ on system performance. The plot reveals a substantial decrease in OP as $\overline{\gamma}_t$ increases. This performance enhancement is expected, as a higher value of $\overline{\gamma}_t$ reduces the effects of fading on the THz link, thereby strengthening this part of the hybrid network. As a result, the system experiences improved data rates and enhanced overall efficiency. Comparative analysis between switching techniques indicate that at  $\overline{\gamma}_t$ = $40$ dB, the OP for the hard switching scheme is approximately $1.51 \times 10^{-3}$. In contrast, for soft switching scheme the OP falls to $7.8 \times 10^{-4}$ under identical SNR conditions. This corresponds to a $48.34\%$ performance improvement, highlighting the superior performance of the soft switching mechanism in high-SNR regimes.
\subsection{Comparative Analysis with Existing Works}

\begin{figure}[!ht]
\vspace{0mm}
\centerline{\includegraphics[width=0.35\textwidth,angle=0]{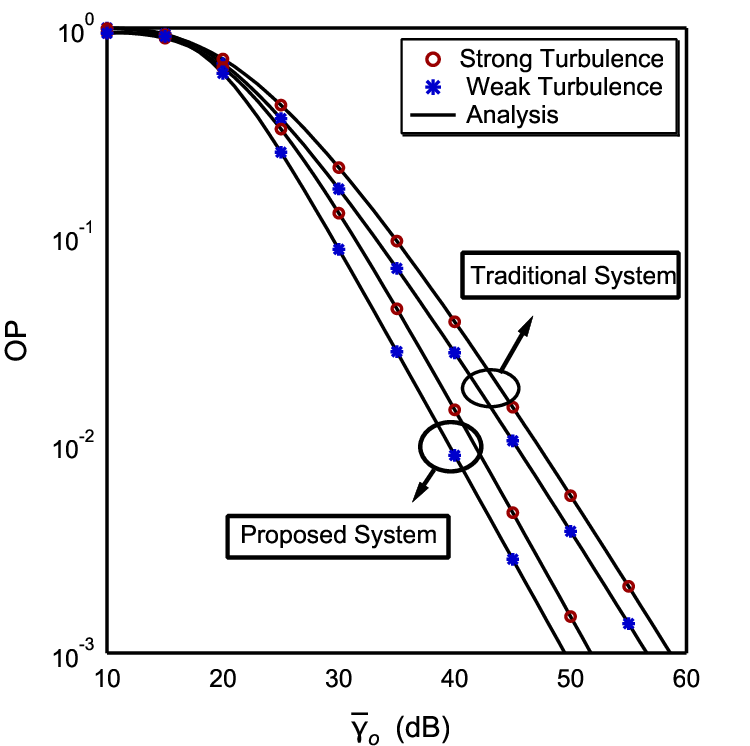}}
    \vspace{0mm}
    \caption{OP versus  \(\overline{\gamma }_o\) for different network systems.}
    \label{fig X}
\end{figure}

Fig. \ref{fig X} presents a comparison with the existing models and proposed model.
Most of the current literature focuses on dual-hop FSO-RF systems. However, a major limitation of these models is their reliability, as FSO links are highly susceptible to blockage, especially in urban environments. To address this issue, we propose a dual-hop mixed network with parallel FSO and THz links in the first hop. As observed in the figure, when $\overline{\gamma }_r = 35$ dB for the hard switching case, the OP for the traditional FSO-RF model is $9.48 \times 10^{-2}$. In contrast, the proposed FSO/THz-RF model achieves a significantly lower OP of $4.47 \times 10^{-2}$ at the same SNR level. This corresponds to an overall improvement of $52.54\%$ in outage performance.
\begin{figure}[!ht]
\vspace{0mm}
\centerline{\includegraphics[width=0.35\textwidth,angle=0]{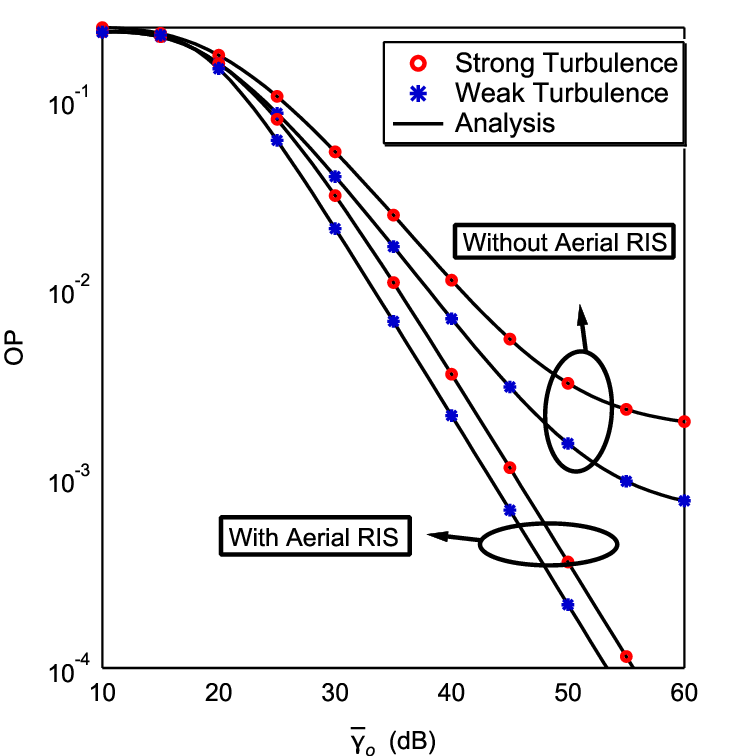}}
    \vspace{0mm}
    \caption{OP versus  \(\overline{\gamma }_o\) under various atmospheric turbulence Conditions. }
    \label{fig YY}
\end{figure}

Fig. \ref{fig YY} illustrates the OP versus \(\overline{\gamma }_o\) under various turbulence scenarios as shown in Table \ref{RIS_comparison}.
\begin{table*}[!ht]
\centering
\caption{OP Comparison with and without Aerial RIS for Various $\overline{\gamma}_r$ and $\overline{\gamma}_o$}
\begin{tabular}{llccl}
\hline
\multicolumn{5}{c}{\textbf{Performance of System Without Aerial RIS vs With Aerial RIS}} \\ \hline
\multicolumn{1}{l}{\multirow{5}{*}{$\overline{\gamma}_r = 10\ \mathrm{dB}$}} &
  \multicolumn{1}{l}{$\overline{\gamma}_o = 10\ \mathrm{dB}$} &
  \multicolumn{1}{c}{$2.50 \times 10^{-1}$} &
  \multicolumn{1}{c}{$2.31 \times 10^{-1}$} &
  $7.60 \%$ \\
\multicolumn{1}{l}{} &
  \multicolumn{1}{l}{$\overline{\gamma}_o = 15\ \mathrm{dB}$} &
  \multicolumn{1}{c}{$2.49 \times 10^{-1}$} &
  \multicolumn{1}{c}{$2.17 \times 10^{-1}$} &
  $12.85 \%$ \\
\multicolumn{1}{l}{} &
  \multicolumn{1}{l}{$\overline{\gamma}_o = 20\ \mathrm{dB}$} &
  \multicolumn{1}{c}{$2.46 \times 10^{-1}$} &
  \multicolumn{1}{c}{$1.57 \times 10^{-1}$} &
  $36.18 \%$ \\
\multicolumn{1}{l}{} &
  \multicolumn{1}{l}{$\overline{\gamma}_o = 25\ \mathrm{dB}$} &
  \multicolumn{1}{c}{$2.43 \times 10^{-1}$} &
  \multicolumn{1}{c}{$9.51 \times 10^{-2}$} &
  $60.86 \%$ \\
\multicolumn{1}{l}{} &
  \multicolumn{1}{l}{$\overline{\gamma}_o = 30\ \mathrm{dB}$} &
  \multicolumn{1}{c}{$2.40 \times 10^{-1}$} &
  \multicolumn{1}{c}{$8.30 \times 10^{-2}$} &
  $65.42 \%$ \\
\hline
\multicolumn{1}{l}{\multirow{5}{*}{$\overline{\gamma}_r = 15\ \mathrm{dB}$}} &
  \multicolumn{1}{l}{$\overline{\gamma}_o = 10\ \mathrm{dB}$} &
  \multicolumn{1}{c}{$2.30 \times 10^{-1}$} &
  \multicolumn{1}{c}{$2.20 \times 10^{-1}$} &
  $4.35 \%$ \\
\multicolumn{1}{l}{} &
  \multicolumn{1}{l}{$\overline{\gamma}_o = 15\ \mathrm{dB}$} &
  \multicolumn{1}{c}{$2.38 \times 10^{-1}$} &
  \multicolumn{1}{c}{$2.13 \times 10^{-1}$} &
  $10.50 \%$ \\
\multicolumn{1}{l}{} &
  \multicolumn{1}{l}{$\overline{\gamma}_o = 20\ \mathrm{dB}$} &
  \multicolumn{1}{c}{$2.12 \times 10^{-1}$} &
  \multicolumn{1}{c}{$1.46 \times 10^{-1}$} &
  $31.13 \%$ \\
\multicolumn{1}{l}{} &
  \multicolumn{1}{l}{$\overline{\gamma}_o = 25\ \mathrm{dB}$} &
  \multicolumn{1}{c}{$1.79 \times 10^{-1}$} &
  \multicolumn{1}{c}{$6.27 \times 10^{-2}$} &
  $64.97 \%$ \\
\multicolumn{1}{l}{} &
  \multicolumn{1}{l}{$\overline{\gamma}_o = 30\ \mathrm{dB}$} &
  \multicolumn{1}{c}{$1.54 \times 10^{-1}$} &
  \multicolumn{1}{c}{$5.39 \times 10^{-2}$} &
  $65.00 \%$ \\
\hline
\multicolumn{1}{l}{\multirow{5}{*}{$\overline{\gamma}_r = 20\ \mathrm{dB}$}} &
  \multicolumn{1}{l}{$\overline{\gamma}_o = 10\ \mathrm{dB}$} &
  \multicolumn{1}{c}{$2.42 \times 10^{-1}$} &
  \multicolumn{1}{c}{$2.30 \times 10^{-1}$} &
  $4.96 \%$ \\
\multicolumn{1}{l}{} &
  \multicolumn{1}{l}{$\overline{\gamma}_o = 15\ \mathrm{dB}$} &
  \multicolumn{1}{c}{$2.26 \times 10^{-1}$} &
  \multicolumn{1}{c}{$2.02 \times 10^{-1}$} &
  $10.62 \%$ \\
\multicolumn{1}{l}{} &
  \multicolumn{1}{l}{$\overline{\gamma}_o = 20\ \mathrm{dB}$} &
  \multicolumn{1}{c}{$1.73 \times 10^{-1}$} &
  \multicolumn{1}{c}{$1.41 \times 10^{-1}$} &
  $18.50 \%$ \\
\multicolumn{1}{l}{} &
  \multicolumn{1}{l}{$\overline{\gamma}_o = 25\ \mathrm{dB}$} &
  \multicolumn{1}{c}{$1.05 \times 10^{-1}$} &
  \multicolumn{1}{c}{$6.83 \times 10^{-2}$} &
  $35.00 \%$ \\
\multicolumn{1}{l}{} &
  \multicolumn{1}{l}{$\overline{\gamma}_o = 30\ \mathrm{dB}$} &
  \multicolumn{1}{c}{$5.35 \times 10^{-2}$} &
  \multicolumn{1}{c}{$1.87 \times 10^{-2}$} &
  $65.04 \%$ \\
\hline
\end{tabular}
\label{RIS_comparison}
\end{table*}
Two different configurations are considered in this analysis: (a) the use of an aerial RIS in the RF link and (b) the absence of RIS in the RF link. The results clearly demonstrate that integrating an aerial RIS into the proposed model leads to a significant performance enhancement. For example, at $30$ dB, the OP for the system without RIS is $5.34 \times 10^{-2}$, whereas the inclusion of an aerial RIS reduces the OP to $3.13 \times 10^{-2}$. This represents an improvement of approximately $41.39\%$ in terms of outage probability.
\subsection{New Findings and Network Design Guidelines}
\label{sec5}
\noindent
Based on the new findings from the investigation into the OP and BER of the proposed system, the following design guidelines should be considered: 
\begin{itemize}
    \item The comparative analysis of detection techniques in Fig. \ref{fig 2} reveals that HD outperforms IM/DD technique. Therefore, network engineers should consider integrating HD methods into their communication systems, especially in scenarios involving long-distance transmissions or adverse environmental conditions, to improve signal quality and system performance.
    
    \item The analysis in Fig. \ref{fig 4} indicates that increasing the number of antennas $N_t$ at the access point results in considerable performance gains. Network engineers are encouraged to implement multiple antennas in their THz systems, which will improve SNR and increase data rates. Moreover, our analysis suggests that a larger receiver aperture should be used to reduce the impact of pointing error.
    
    \item Fig. \ref{fig X} highlights a decrease in OP from $9.48 \times 10^{-2}$ for traditional systems to $4.47 \times 10^{-2}$ for the proposed hybrid system at $35$ dB. This $52.54\% $ enhancement suggests that the integration of parallel FSO/THz technologies in network designs can increase the reliability and improve overall operational performance.
    
    \item As shown in Fig. \ref{fig 11}, the system performance of the proposed model deteriorates with increasing environmental clutter. To mitigate the adverse effects of dense environments, it is recommended to deploy RIS with a greater number of reflecting elements. The integration of RIS with additional elements significantly enhances overall system performance.
    
    \item Fig. \ref{fig 8} highlights the critical role of UAV positioning in optimizing system performance. Placing the UAV directly above the midpoint between the access point and the mobile user significantly reduces the ABER. The results demonstrate that symmetric UAV placement leads to a considerably lower ABER compared to asymmetric positioning. Therefore, the proposed model recommends that network engineers position the aerial RIS at the midpoint between the source and the receiver for optimal performance.
    
    \item Fig. \ref{fig 9} illustrates the impact of UAV altitude on OP, revealing that as the UAV altitude increases, the OP also rises, leading to degraded system performance. For optimal communication, it is essential to maintain a balance between altitude and LoS conditions. Numerical results indicate that lowering the UAV altitude significantly improves OP. Therefore, the proposed model also recommends the use of soft switching strategies to help mitigate the performance degradation.
\end{itemize}
\section{Conclusions}
\label{sec6}
This paper investigated a mixed dual-hop relay-assisted wireless communication system, where the first hop employed hybrid FSO/THz links and the second hop utilized an aerial RIS-assisted network. Closed-form expressions for OP and ABER were derived for both hard and soft switching schemes and validated through extensive Monte Carlo simulations. The results demonstrated that soft switching consistently outperformed hard switching by adapting to varying link conditions, while the HD technique achieved better performance than IM/DD. Severe atmospheric turbulence and high pointing errors notably degraded system performance; however, these effects were effectively mitigated by deploying an aerial RIS with multiple reflecting elements. Moreover, the findings underscored the critical role of UAV placement, showing that optimal positioning and altitude of the RIS significantly enhanced system reliability. The impact of modulation schemes was also analyzed, revealing that lower-order schemes such as $4$-QAM offered better error performance than higher-order schemes like $64$-QAM under hybrid channel conditions. Finally, the asymptotic analysis indicated that the diversity order was governed by factors such as fading parameters, THz link antenna diversity, RF link shaping parameters, atmospheric turbulence, pointing errors, and the detection technique used in the FSO link.
\bibliographystyle{IEEEtran}
\bibliography{IEEEabrv,main}

\begin{thebibliography}{10}
\providecommand{\url}[1]{#1}
\csname url@samestyle\endcsname
\providecommand{\newblock}{\relax}
\providecommand{\bibinfo}[2]{#2}
\providecommand{\BIBentrySTDinterwordspacing}{\spaceskip=0pt\relax}
\providecommand{\BIBentryALTinterwordstretchfactor}{4}
\providecommand{\BIBentryALTinterwordspacing}{\spaceskip=\fontdimen2\font plus
\BIBentryALTinterwordstretchfactor\fontdimen3\font minus \fontdimen4\font\relax}
\providecommand{\BIBforeignlanguage}[2]{{%
\expandafter\ifx\csname l@#1\endcsname\relax
\typeout{** WARNING: IEEEtran.bst: No hyphenation pattern has been}%
\typeout{** loaded for the language `#1'. Using the pattern for}%
\typeout{** the default language instead.}%
\else
\language=\csname l@#1\endcsname
\fi
#2}}
\providecommand{\BIBdecl}{\relax}
\BIBdecl

\bibitem{r1}
V.~K. Chapala and S.~M. Zafaruddin, ``Unified performance analysis of reconfigurable intelligent surface empowered free-space optical communications,'' \emph{IEEE Transactions on Communications}, vol.~70, no.~4, pp. 2575--2592, 2021.

\bibitem{li2023ris}
X.~Li, Y.~Li, X.~Song, L.~Shao, and H.~Li, ``{RIS} assisted {UAV} for weather-dependent satellite terrestrial integrated network with hybrid {FSO/RF} systems,'' \emph{IEEE Photonics Journal}, 2023.

\bibitem{n1}
A.~Douik, H.~Dahrouj, T.~Y. Al-Naffouri, and M.-S. Alouini, ``Hybrid radio/free-space optical design for next generation backhaul systems,'' \emph{IEEE Transactions on Communications}, vol.~64, no.~6, pp. 2563--2577, 2016.

\bibitem{n2}
M.~Najafi, V.~Jamali, and R.~Schober, ``Optimal relay selection for the parallel hybrid {RF/FSO} relay channel: Non-buffer-aided and buffer-aided designs,'' \emph{IEEE Transactions on Communications}, vol.~65, no.~7, pp. 2794--2810, 2017.

\bibitem{n3}
A.~Touati, A.~Abdaoui, F.~Touati, M.~Uysal, and A.~Bouallegue, ``On the effects of combined atmospheric fading and misalignment on the hybrid {FSO/RF} transmission,'' \emph{Journal of Optical Communications and Networking}, vol.~8, no.~10, pp. 715--725, 2016.

\bibitem{n22}
B.~Bag, A.~Das, I.~S. Ansari, A.~Proke{\v{s}}, C.~Bose, and A.~Chandra, ``Performance analysis of hybrid {FSO} systems using {FSO/RF-FSO} link adaptation,'' \emph{IEEE Photonics Journal}, vol.~10, no.~3, pp. 1--17, 2018.

\bibitem{n23}
M.~Usman, H.-C. Yang, and M.-S. Alouini, ``Practical switching-based hybrid {FSO/RF} transmission and its performance analysis,'' \emph{IEEE Photonics journal}, vol.~6, no.~5, pp. 1--13, 2014.

\bibitem{n4}
S.~Sharma and A.~Madhukumar, ``Switching-based cooperative decode-and-forward relaying for hybrid {FSO/RF} networks,'' \emph{Journal of Optical Communications and Networking}, vol.~11, no.~6, pp. 267--281, 2019.

\bibitem{n5}
M.~P. Ninos, P.~Mukherjee, C.~Psomas, and I.~Krikidis, ``Full-duplex {DF relaying with parallel hybrid FSO/RF} transmissions,'' \emph{IEEE Open Journal of the Communications Society}, vol.~2, pp. 2502--2515, 2021.

\bibitem{r11}
W.~A. Alathwary and E.~S. Altubaishi, ``On the performance analysis of decode-and-forward multi-hop hybrid {FSO/RF} systems with hard-switching configuration,'' \emph{IEEE Photonics Journal}, vol.~11, no.~6, pp. 1--12, 2019.

\bibitem{n6}
Y.~F. Al-Eryani, A.~M. Salhab, S.~A. Zummo, and M.-S. Alouini, ``Protocol design and performance analysis of multiuser mixed {RF} and hybrid {FSO/RF} relaying with buffers,'' \emph{Journal of Optical Communications and Networking}, vol.~10, no.~4, pp. 309--321, 2018.

\bibitem{n7}
B.~Bag, A.~Das, C.~Bose, and A.~Chandra, ``Improving the performance of a {DF} relay-aided {FSO} system with an additional source--relay mmwave {RF} backup,'' \emph{Journal of Optical Communications and Networking}, vol.~12, no.~12, pp. 390--402, 2020.

\bibitem{n8}
M.~Z. Hassan, M.~J. Hossain, J.~Cheng, and V.~C. Leung, ``Hybrid {RF/FSO} backhaul networks with statistical-{QoS}-aware buffer-aided relaying,'' \emph{IEEE Transactions on Wireless Communications}, vol.~19, no.~3, pp. 1464--1483, 2019.

\bibitem{thz}
P.~K. Singya, B.~Makki, A.~D’Errico, and M.-S. Alouini, ``Hybrid {FSO/THz}-based backhaul network for mm{W}ave terrestrial communication,'' \emph{IEEE Transactions on Wireless Communications}, 2022.

\bibitem{r6}
D.~Singh and R.~Swaminathan, ``Comprehensive performance analysis of hovering {UAV-based FSO} communication system,'' \emph{IEEE Photonics Journal}, vol.~14, no.~5, pp. 1--13, 2022.

\bibitem{r13}
R.-R. Lu, J.-Y. Wang, X.-T. Fu, S.-H. Lin, Q.~Wang, and B.~Zhang, ``Performance analysis and optimization for {UAV-based FSO} communication systems,'' \emph{Physical Communication}, vol.~51, p. 101594, 2022.

\bibitem{r14}
J.-Y. Wang, Y.~Ma, R.-R. Lu, J.-B. Wang, M.~Lin, and J.~Cheng, ``Hovering {UAV-based FSO communications: C}hannel modelling, performance analysis, and parameter optimization,'' \emph{IEEE Journal on Selected Areas in Communications}, vol.~39, no.~10, pp. 2946--2959, 2021.

\bibitem{r2}
P.~K. Singya and M.-S. Alouini, ``Performance of {UAV-assisted multiuser terrestrial-satellite communication system over mixed FSO/RF} channels,'' \emph{IEEE Transactions on Aerospace and Electronic Systems}, vol.~58, no.~2, pp. 781--796, 2021.

\bibitem{n26}
L.~Qu, G.~Xu, Z.~Zeng, N.~Zhang, and Q.~Zhang, ``{UAV-assisted RF/FSO} relay system for space-air-ground integrated network: A performance analysis,'' \emph{IEEE Transactions on Wireless Communications}, vol.~21, no.~8, pp. 6211--6225, 2022.

\bibitem{n29}
G.~Xu and Z.~Song, ``Performance analysis of a {UAV-assisted RF/FSO} relaying systems for internet of vehicles,'' \emph{IEEE Internet of Things Journal}, vol.~9, no.~8, pp. 5730--5741, 2021.

\bibitem{n30}
G.~Xu, S.~Lu, L.~Qu, Q.~Zhang, Z.~Song, and B.~Ai, ``Outage probability and average {BER of UAV-Assisted RF/FSO} system for space-air-ground integrated networks under angle-of-arrival fluctuations,'' \emph{IEEE Internet of Things Journal}, 2024.

\bibitem{n25}
G.~Xu, N.~Zhang, M.~Xu, Z.~Xu, Q.~Zhang, and Z.~Song, ``Outage probability and average {BER of UAV-assisted dual-hop FSO} communication with amplify-and-forward relaying,'' \emph{IEEE Transactions on Vehicular Technology}, vol.~72, no.~7, pp. 8287--8302, 2023.

\bibitem{r8}
M.~T. Dabiri, M.~Hasna, S.~Althunibat, and K.~Qaraqe, ``{UAV-Based Dynamic FSO} access networks: Technological comparison, design considerations, and future directions,'' \emph{IEEE Wireless Communications}, vol.~32, no.~2, pp. 247--253, 2025.

\bibitem{r9}
D.~Kim, H.~S. Park, and B.~C. Jung, ``Performance analysis of {RIS-assisted dual-hop mixed FSO-RF UAV} communication systems,'' \emph{Digital Communications and Networks}, 2025.

\bibitem{n17}
I.~S. Ansari, M.~M. Abdallah, M.-S. Alouini, and K.~A. Qaraqe, ``A performance study of two hop transmission in mixed underlay {RF and FSO} fading channels,'' in \emph{2014 IEEE Wireless Communications and Networking Conference (WCNC)}.\hskip 1em plus 0.5em minus 0.4em\relax IEEE, 2014, pp. 388--393.

\bibitem{n9}
E.~Zedini, H.~Soury, and M.-S. Alouini, ``On the performance analysis of dual-hop mixed {FSO/RF} systems,'' \emph{IEEE Transactions on Wireless Communications}, vol.~15, no.~5, pp. 3679--3689, 2016.

\bibitem{n10}
Q.~Sun, Z.~Zhang, Y.~Zhang, M.~L{\'o}pez-Ben{\'\i}tez, and J.~Zhang, ``Performance analysis of dual-hop wireless systems over mixed {FSO/RF} fading channel,'' \emph{IEEE Access}, vol.~9, pp. 85\,529--85\,542, 2021.

\bibitem{n16}
O.~M.~S. Al-Ebraheemy, A.~M. Salhab, A.~Chaaban, S.~A. Zummo, and M.-S. Alouini, ``Precise performance analysis of dual-hop mixed {RF/unified-FSO DF relaying with heterodyne detection and two IM-DD} channel models,'' \emph{IEEE Photonics Journal}, vol.~11, no.~1, pp. 1--22, 2019.

\bibitem{n11}
Z.~Zhang, Q.~Sun, M.~López-Benítez, X.~Chen, and J.~Zhang, ``Performance analysis of dual-hop {RF/FSO} relaying systems with imperfect {CSI},'' \emph{IEEE Transactions on Vehicular Technology}, vol.~71, no.~5, pp. 4965--4976, 2022.

\bibitem{n18}
S.~Anees, P.~Meerur, and M.~R. Bhatnagar, ``Performance analysis of a {DF based dual hop mixed RF-FSO system with a direct RF} link,'' in \emph{2015 IEEE Global Conference on Signal and Information Processing (GlobalSIP)}.\hskip 1em plus 0.5em minus 0.4em\relax IEEE, 2015, pp. 1332--1336.

\bibitem{n19}
E.~Zedini, I.~S. Ansari, and M.-S. Alouini, ``Unified performance analysis of mixed line of sight {RF-FSO} fixed gain dual-hop transmission systems,'' in \emph{2015 IEEE Wireless Communications and Networking Conference (WCNC)}.\hskip 1em plus 0.5em minus 0.4em\relax IEEE, 2015, pp. 46--51.

\bibitem{n12}
G.~Xu and Z.~Song, ``Performance analysis for mixed $\kappa$-$\mu$ fading and {$M$}-distribution dual-hop radio frequency/free space optical communication systems,'' \emph{IEEE Transactions on Wireless Communications}, vol.~20, no.~3, pp. 1517--1528, 2020.

\bibitem{n13}
B.~Ashrafzadeh, E.~Soleimani-Nasab, M.~Kamandar, and M.~Uysal, ``A framework on the performance analysis of dual-hop mixed {FSO-RF} cooperative systems,'' \emph{IEEE Transactions on Communications}, vol.~67, no.~7, pp. 4939--4954, 2019.

\bibitem{n20}
E.~Soleimani-Nasab and M.~Uysal, ``Generalized performance analysis of mixed {RF/FSO} cooperative systems,'' \emph{IEEE Transactions on Wireless Communications}, vol.~15, no.~1, pp. 714--727, 2015.

\bibitem{n14}
R.~Deka, V.~Mishra, I.~Ahmed, S.~Anees, and M.~S. Alam, ``On the performance and optimization of {HAPS assisted dual-hop hybrid RF/FSO} system,'' \emph{IEEE Access}, vol.~10, pp. 80\,976--80\,988, 2022.

\bibitem{n21}
E.~Zedini, I.~S. Ansari, and M.-S. Alouini, ``Performance analysis of mixed {Nakagami-$m$ and Gamma--Gamma dual-hop FSO} transmission systems,'' \emph{IEEE Photonics Journal}, vol.~7, no.~1, pp. 1--20, 2014.

\bibitem{n15}
P.~K. Singya, N.~Kumar, V.~Bhatia, and M.-S. Alouini, ``On the performance analysis of higher order {QAM} schemes over mixed {RF/FSO} systems,'' \emph{IEEE Transactions on Vehicular Technology}, vol.~69, no.~7, pp. 7366--7378, 2020.

\bibitem{r15}
M.~M. Rahman, A.~Badrudduza, N.~A. Sarker, M.~Ibrahim, I.~S. Ansari, and H.~Yu, ``{RIS-aided mixed RF-FSO} wireless networks: Secrecy performance analysis with simultaneous eavesdropping,'' \emph{IEEE Access}, vol.~11, pp. 126\,507--126\,523, 2023.

\bibitem{r17}
A.~M. Salhab and L.~Yang, ``Mixed {RF/FSO relay networks: RIS-equipped RF source vs RIS-aided RF} source,'' \emph{IEEE Wireless Communications Letters}, vol.~10, no.~8, pp. 1712--1716, 2021.

\bibitem{r18}
A.~Sikri, A.~Mathur, and G.~Kaddoum, ``Signal space diversity-based distributed {RIS-aided dual-hop mixed RF-FSO} systems,'' \emph{IEEE Communications Letters}, vol.~26, no.~5, pp. 1066--1070, 2022.

\bibitem{odeyemi2022performance}
K.~O. Odeyemi, P.~A. Owolawi, and O.~O. Olakanmi, ``On the performance of reconfigurable intelligent surface in cooperative decode-and-forward relaying for hybrid rf/fso systems.'' \emph{Progress In Electromagnetics Research M}, vol. 110, 2022.

\bibitem{vishwakarma2024ris}
N.~Vishwakarma, R.~Swaminathan, R.~Premanand, S.~Sharma, and A.~Madhukumar, ``Ris-assisted hybrid fso/thz system with diversity combining schemes: A performance analysis,'' \emph{IEEE Internet of Things Journal}, vol.~11, no.~17, pp. 28\,605--28\,622, 2024.

\bibitem{sharma2022performance}
S.~Sharma, N.~Vishwakarma, and R.~Swaminathan, ``Performance analysis of irs-assisted hybrid fso/rf communication system,'' in \emph{2022 National Conference on Communications (NCC)}.\hskip 1em plus 0.5em minus 0.4em\relax IEEE, 2022, pp. 268--273.

\bibitem{ris}
L.~Yang, P.~Li, F.~Meng, and S.~Yu, ``Performance analysis of {RIS-assisted UAV} communication systems,'' \emph{IEEE Transactions on Vehicular Technology}, vol.~71, no.~8, pp. 9078--9082, 2022.

\bibitem{mahmoud2021intelligent}
A.~Mahmoud, S.~Muhaidat, P.~C. Sofotasios, I.~Abualhaol, O.~A. Dobre, and H.~Yanikomeroglu, ``Intelligent reflecting surfaces assisted uav communications for iot networks: Performance analysis,'' \emph{IEEE Transactions on Green Communications and Networking}, vol.~5, no.~3, pp. 1029--1040, 2021.

\bibitem{gong2020toward}
S.~Gong, X.~Lu, D.~T. Hoang, D.~Niyato, L.~Shu, D.~I. Kim, and Y.-C. Liang, ``Toward smart wireless communications via intelligent reflecting surfaces: {A} contemporary survey,'' \emph{IEEE Communications Surveys \& Tutorials}, vol.~22, no.~4, pp. 2283--2314, 2020.

\bibitem{ibrahim}
M.~Ibrahim, A.~Badrudduza, M.~S. Hossen, M.~K. Kundu, I.~S. Ansari, and I.~Ahmed, ``On effective secrecy throughput of underlay spectrum sharing $\alpha$-$\mu$ / ${M}\acute{a}laga$ hybrid model under interference-and-transmit power constraints,'' \emph{IEEE Photonics Journal}, vol.~15, no.~2, pp. 1--13, 2023.

\bibitem{10697101}
A.~B. Sarawar, A.~S.~M. Badrudduza, M.~Ibrahim, I.~S. Ansari, and H.~Yu, ``Secrecy performance analysis of integrated rf-uowc iot networks enabled by uav and underwater ris,'' \emph{IEEE Internet of Things Journal}, vol.~12, no.~3, pp. 2592--2608, 2025.

\bibitem{roisul}
M.~R.~A. Ruku, M.~Ibrahim, A.~Badrudduza, I.~S. Ansari, W.~Khalid, and H.~Yu, ``Effects of co-channel interference on {RIS} empowered wireless networks amid multiple eavesdropping attempts,'' \emph{ICT Express}, vol.~10, no.~3, pp. 491--497, 2024.

\bibitem{rakib}
M.~A. Rakib, M.~Ibrahim, A.~Badrudduza, I.~S. Ansari, S.~Chakravarty, I.~Ahmed, and S.~A. Razzak, ``A {RIS empowered THz-UWO} relay system for air-to-underwater mixed network: Performance analysis with pointing errors,'' \emph{IEEE Internet of Things Journal}, 2024.

\bibitem{shimamoto2006channel}
S.~Shimamoto \emph{et~al.}, ``Channel characterization and performance evaluation of mobile communication employing stratospheric platforms,'' \emph{IEICE transactions on communications}, vol.~89, no.~3, pp. 937--944, 2006.

\bibitem{edition2007table}
S.~Edition, ``Table of integrals, series, and products,'' 2007.

\bibitem{lei2019secure}
H.~Lei, D.~Wang, K.-H. Park, I.~S. Ansari, G.~Pan, and M.-S. Alouini, ``On secure uav communication systems with randomly located eavesdroppers,'' in \emph{2019 IEEE/CIC International Conference on Communications in China (ICCC)}.\hskip 1em plus 0.5em minus 0.4em\relax IEEE, 2019, pp. 201--206.

\bibitem{zedini2020performance}
E.~Zedini, A.~Kammoun, and M.-S. Alouini, ``Performance of multibeam very high throughput satellite systems based on {FSO feeder links with HPA} nonlinearity,'' \emph{IEEE Transactions on Wireless Communications}, vol.~19, no.~9, pp. 5908--5923, 2020.

\bibitem{prudnikov1986integrals}
A.~P. Prudnikov, I.~A. Brychkov, and O.~I. Marichev, \emph{Integrals and series: special functions}.\hskip 1em plus 0.5em minus 0.4em\relax CRC press, 1986, vol.~3.

\bibitem{yang2016engineering}
X.-S. Yang, \emph{Engineering mathematics with examples and applications}.\hskip 1em plus 0.5em minus 0.4em\relax Academic Press, 2016.

\bibitem{ansari2015performance}
I.~S. Ansari, F.~Yilmaz, and M.-S. Alouini, ``Performance analysis of free-space optical links over {M}{\'a}laga (\(\mathcal{M}\)) turbulence channels with pointing errors,'' \emph{IEEE Transactions on Wireless Communications}, vol.~15, no.~1, pp. 91--102, 2015.

\end{thebibliography}

\end{document}